\documentclass[11pt,twoside]{book} 
\usepackage{asp2008n}
\usepackage{times}
\usepackage{lscape}
\usepackage{epsf}

\usepackage{graphicx}
\usepackage{amssymb}
\usepackage{longtable}
\usepackage[figuresright]{rotating}
\usepackage{multirow}

\newcommand{\simless}{\mathbin{\lower 3pt\hbox {$\rlap{\raise 5pt\hbox{$\char'074$}}\mathchar"7218$}}}

\mainmatter

\newlength{\deftabcolsep}
\begin{document}

\title{Star Forming Regions in Cassiopeia}
\author{M\'aria Kun}
\affil{Konkoly Observatory, H-1525 Budapest, P.O. Box 67, Hungary}

\begin{abstract}
This chapter describes the Galactic star forming regions in the
constellation Cassiopeia, in the Galactic coordinate range $120\deg
\la l \la 130\deg$, $-5\deg \la b \la 15\deg$. At $|b| > 10\deg$ the
nearby clouds L\,1333 and L\,1340 are found in this region. The local
arm of the Galaxy in Cassiopeia contains only a few star forming
regions, smaller and less active than the OB~associations of the
neighboring Cepheus. Five members of this system, LkH$\alpha$\,198 and
its environment, L\,1287, L\,1293, L\,1302/NGC\,255, and S\,187 are
discussed.  Several more distant OB~associations and giant star
forming regions in Cassiopeia are associated with the Perseus arm at
2.0--3.0\,kpc. Among these, the Herbig~Be star MWC\,1080 is discussed in
this chapter.
\end{abstract}

\section{Overview}

\begin{figure*}[ht!]
\centerline{
\includegraphics[width=5.25in,draft=False]{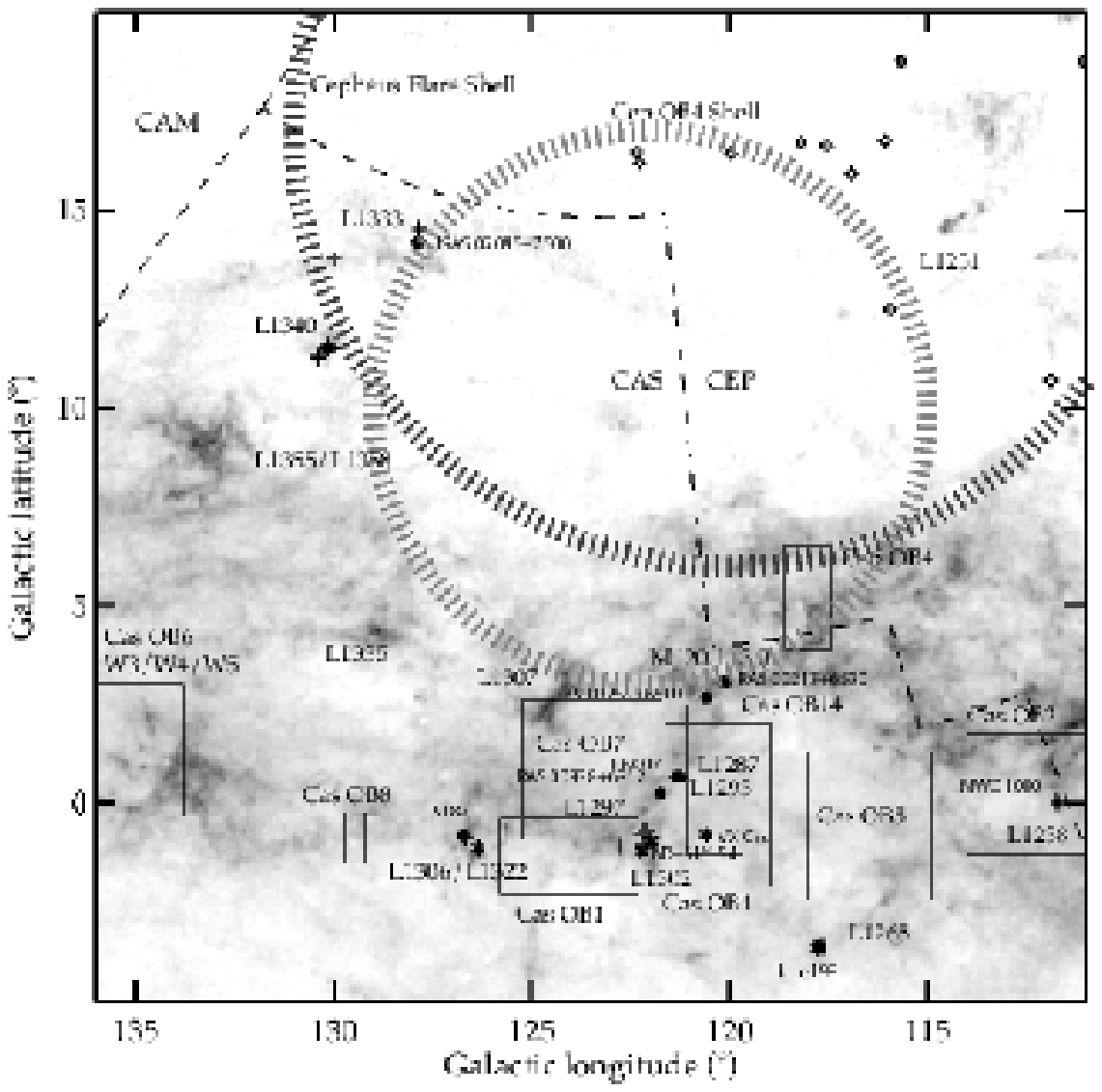}
}
\caption{Distribution of visual extinction in the Cassiopeia region
\citep[adopted from][]{DUK}. The prominent clouds and the young stellar objects
associated with the clouds discussed in this
chapter are marked. Filled circles are outflow sources, plusses are classical
T Tauri stars, asterisks mark Herbig Ae/Be stars, and diamonds weak-line T Tauri
stars. Rectangles mark the nominal boundaries of the OB associations \citep{Humphreys}.
Dashed lines indicate the boundaries of the constellation. Large circles drawn
by radial dashes show the approximate positions of the Cepheus Flare Shell and
Cep~OB4 Shell \citep{Olano}.}
\label{Fig1}
\end{figure*}

The large-scale distribution of the dark clouds in Cassiopeia, adopted from the
Atlas and Catalog of Dark Clouds \citep{DUK}, is displayed in Fig.~\ref{Fig1}. The most prominent
clouds and young stellar objects are marked.

The high Galactic latitude region of the constellation Cassiopeia contains
a few nearby star forming molecular clouds. The nearest cloud
is Lynds~1333 at ($l,b$)=(129\deg,\linebreak +15\deg) and at
$d \approx 180$\,pc,
associated with a few low-mass pre-main sequence stars. Lynds~1340, a region
of intermediate and low mass star formation, can be found at\linebreak
($l,b$)=(130\deg,+11\fdg5) at
$d \approx 600$\,pc. Towards lower latitudes, the cloud
complex\linebreak  L\,1355/L\,1358 is
located at ($l,b$) $\sim$ (133\fdg5,+8\fdg6).
This complex is illuminated by a loose group of B and A stars and the classical
Cepheid SU~Cas \citep{Turner}. The average distance of these stars, determined by
\citet{Turner} is $d = 258 \pm 3$\,pc. No star formation has been
observed in these clouds. The starless cores of L\,1355 and L\,1358 have been targets of
several studies aimed at observing initial conditions of low-mass star formation
and protostellar infall \citep*{Lee99,Lee01,Park04}.

Star forming regions at various distances can be found at latitudes ($|b| < 10\deg$).
The catalog of CO clouds associated with IRAS point sources, published
by \citet{Kerton}, contains numerous likely star forming regions which deserve follow-up studies.
L\,1265, L\,1287, and S\,187, located between 600--1000\,pc, are associated with the Orion
arm of our Galaxy. The most distant star forming regions,
associated with the Perseus arm, include several OB associations \citep[e.g.][]{Humphreys,GS92}
located between 1.8--2.8\,kpc, and  the Herbig~Be star MWC\,1080, one
of the best studied members of this class of objects.
The most prominent member of this system, the giant star forming complex W3/W4/W5 is
discussed in the chapter by Megeath et~al.
The regions discussed in the present chapter are  listed in Table~\ref{Tab_cloud}.

\begin{table}
\caption{Star-forming molecular clouds in Cassiopeia}
\label{Tab_cloud}
\smallskip
\begin{center}
{\footnotesize
\begin{tabular}{llcl}
\tableline
Dark Cloud$^a$ & CO cloud$^b$~  & d\,(pc) & Associated YSO \\
\noalign{\smallskip}
\tableline
\noalign{\smallskip}
L\,1238 & 111.7+00.0 & 2200 & MWC\,1080 \\
L\,1265 & 117.8$-$03.6  & 600  &  LkH$\alpha$ 198 \\
$\cdots$ &  120.1+03.0$^c$  & 850 & IRAS 00213+6530, IRAS 00259+6510 \\
L\,1287, TDS463 & 121.3+00.6 & 	850 & RNO\,1, IRAS 00338+6312 \\
L\,1293 & 121.7+00.2 & 850 & IRAS 00379+6248 \\
L\,1302 & 122.0$-$01.1 & 600 & BD\,+61\deg154, LkH$\alpha$201 \\
L\,1317 & 126.6$-$00.6 & 800 & S\,187  \\
L\,1333 & 129.0+13.8 & 180 & IRAS 02086+7600 \\
L\,1340 & 131.1+11.6 & 600 & RNO\,7,8,9 \\
\noalign{\smallskip}
\tableline
\noalign{\smallskip}
\multicolumn{4}{l}{\parbox{0.8\textwidth}{\footnotesize
    $^a$ \citet{Lynds,TDS};  }}\\[1ex]
\multicolumn{4}{l}{\parbox{0.8\textwidth}{\footnotesize
    $^b$\citet{Yonekura}; }}\\[1ex]
\multicolumn{4}{l}{\parbox{0.8\textwidth}{\footnotesize
    $^c$This cloud lies within the region studied in this chapter,
    but, as  it is associated with Cep~OB\,4 in the literature, it is
    discussed in the Cepheus chapter.}}\\

\end{tabular}}
\end{center}
\smallskip
\end{table}

\section{Lynds 1333}

\begin{figure*}[h!]
\centerline{
\includegraphics[width=5.25in]{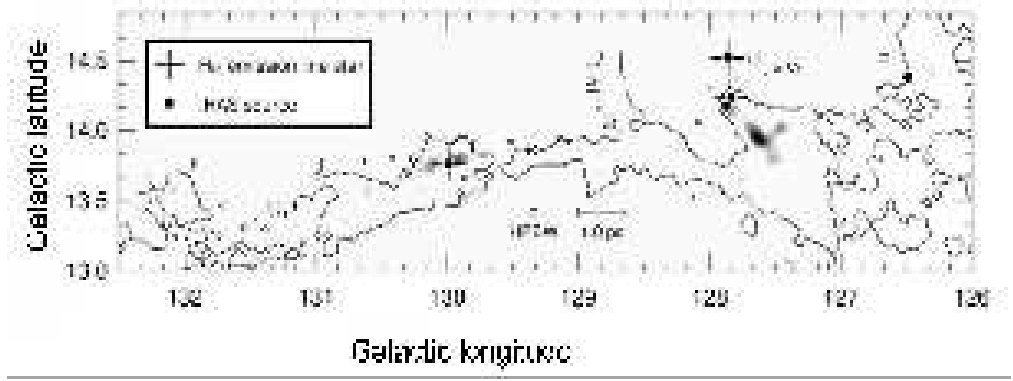}
}
\caption{Integrated intensity map of the $^{13}$CO (contour) and C$^{18}$O (gray scale)
and the positions of the bona fide YSOs in the L\,1333 cloud complex.
Crosses mark the OKS\,H$\alpha$ stars \citep{Obayashi}, and dots are IRAS point sources.}
\label{Fig_1333}
\end{figure*}

Lynds~1333 is a small dark cloud of opacity class~6
at  ($l,b$)=($128\fdg88,+13\fdg71$) \citep{Lynds}. \citet{Obayashi} derived a distance of
180$\pm20$\,pc using a Wolf diagram, and mapped the cloud in
$^{13}$CO and C$^{18}$O. According to the available data, the L\,1333
dark cloud itself is starless, and has been included in several studies of
starless cores \citep[e.g.][]{Lee99,Lee01,Lee04}. $^{13}$CO and C$^{18}$O
observations by \citet{Obayashi} have shown this dark cloud to be part of a long
(some 30\,pc), filamentary cloud complex ($126\deg \leq l \leq 133\deg,
+13\deg \la b \la +15\deg$). They refer to this molecular
complex as the {\em L1333 molecular cloud\/} (Fig.~\ref{Fig_1333}).
Star formation in the L\,1333 molecular
cloud has been indicated by the presence of the protostellar-like (Class~I SED)
source IRAS~02086+7600 and by three other IRAS sources
coinciding with H$\alpha$ emission stars. Comparison of the properties of dense
C$^{18}$O cores of L\,1333 with those of other nearby star forming clouds
(Taurus, Ophiuchus, Lupus, and Chamaeleon) suggests that L\,1333 is among the
smallest known star forming molecular cloud complexes
\citep{Onishi99,Tachi00, Tachi02}.
It represents an extreme environment of star formation, where a few
low-mass stars are being formed in a small, filamentary cloud complex.

The total mass of $^{13}$CO molecular clouds shown in Fig.~\ref{Fig_1333}
is estimated to be about 720\,$M_{\sun}$.
Thirteen  C$^{18}$O cores, which are characterized by mean mass of
9\,$M_{\sun}$ and  mean density of
$1.4\times10^{4}$\,cm$^{-3}$, are embedded in the $^{13}$CO cloud.

Eighteen H$\alpha$ emission line stars have been detected
within or near the $^{13}$CO clouds on objective prism plates.
Spectroscopic follow-up observations confirmed the pre-main sequence nature
of only three of them. One of the three, OKS98\,H$\alpha$\,6, proved to be a
visual double with two T~Tauri type components, separated by 1\farcs8.
Table~\ref{Tab_1333} shows the IRAS and 2MASS
data for these objects, as well as spectral types and optical
photometric results \citep{Kun06}. Their distribution with
respect to the molecular cloud
is displayed in Fig.~\ref{Fig_1333}.  The source IRAS 02086+7600
is associated with a C$^{18}$O core and exhibits a Class~I type infrared
spectrum, yet coincides with a faint visible star.

There are several weak-line T Tauri stars in the cloudless region between
L\,1333 and the high-longitude edge of the Cepheus flare ($116\deg < l < 124\deg,
b \sim 17\deg$, found by \citet{Tachi05}.
Their relation to the cloud complex is unknown.

\citet*{Olano} studied the space distribution and kinematics of the interstellar matter
in the Cepheus flare and in the neighboring Cassiopeia clouds, using the Leiden--Dwingeloo
HI data and the Columbia Survey CO data. They found that the broad and
often double-peaked spectral line profiles suggest that the Cepheus Flare
and Cassiopeia clouds form a big expanding shell that encloses an old supernova remnant.
Assuming a distance of 300~pc for the center
of the shell they derived a radius of approximately 50~pc, expansion velocity of
4~km\,s$^{-1}$, and HI mass of $1.3 \times 10^{4}$~M$_{\sun}$ for the Cepheus Flare Shell.
L\,1333 probably lies on the approaching side of the Cepheus Flare Shell, whose approximate
position is plotted in Fig.~\ref{Fig1}.

\begin{table}[ht!]
\caption{Low-mass pre-main sequence stars in L\,1333}
\label{Tab_1333}
\smallskip
\begin{center}
{\footnotesize
\begin{tabular}{lrrrr}
\tableline
\noalign{\smallskip}
IRAS   & $F_{12}$ & $F_{25}$ & $F_{60}$ & $F_{100}$ \\
2MASS  & & $J$ & $H$ & $K_s$ \\
OKS98 H$\alpha$  & Sp. & $V$ & $R_C$ & $I_C$ \\
\noalign{\smallskip}
\tableline
\noalign{\smallskip}
IRAS 02086+7600  & 0.25 & 2.68 & 7.41 & 10.49 \\
02134360+7615059  && 13.715 & 12.254  & 11.193 \\
 $\cdots$ & $\cdots$ & 19.71 & 17.81 & 16.47 \\[5pt]
IRAS F02084+7605  & 0.12  & 0.14 & $<$0.29 & $<$7.18 \\
02133179+7619127 && 10.648 & 9.725  & 9.317 \\
OKS98 H$\alpha$\,5 & M0.5IV & 15.49 & 14.12 & 12.76 \\[5pt]
IRAS 02103+7621   &  0.51 & 0.58 & 0.33: & $<$1.93 \\
02152532+7635196 & & 10.222 & 9.342  & 8.701  \\
OKS98 H$\alpha$\,6\,N  & K7V & 13.19 & 12.57 & 11.78 \\
OKS98 H$\alpha$\,6\,S  & M2IV & $\cdots$ & 15.73 & 14.10 \\[5pt]
IRAS 02368+7453 &  0.44 & 0.89 & 0.89 & 3.21 \\
02420054+7505473  & & 10.920 & 9.654 & 8.963 \\
OKS98 H$\alpha$\,16   & K5III  & 17.42 & 15.57 & 13.52 \\
\noalign{\smallskip}
\tableline
\end{tabular}}
\end{center}
\end{table}

\section{Lynds 1340}
Lynds~1340 is a dark cloud of opacity class 5 at ($l,b$)=(130\deg, +11\deg)
 \citep{Lynds}.
It is associated with the reflection nebula DG~9 \citep{DG},
illuminated by B and A-type stars. The small nebulosities
RNO~7, 8, and 9 \citep{RNO}, associated with
the cloud, are probably signposts of recent star formation.
The basic properties of the cloud were studied by \citet{KOS94}.
They presented $^{13}$CO and C$^{18}$O maps of the cloud, and obtained
a distance of 600$\pm$60\,pc by averaging the results obtained from a
Wolf diagram, as well as the spectral classification and
\ubv photometry of the stars illuminating DG\,9.
They also provided a list of candidate young stellar objects, selected from
the IRAS catalogs and H$\alpha$ emission stars identified on objective prism
Schmidt plates.
\citet{Yonekura} found that L\,1340 contains some 1300 $M_{\sun}$ of molecular mass.

The distribution of $^{13}$CO has revealed  three clumps within
the cloud, denoted as L\,1340\,A, B, and C by \citet{KOS94}. Each clump is
associated with massive C$^{18}$O cores, and with a number of IRAS point sources
and H$\alpha$ emission stars.

The densest regions of L\,1340 were mapped in the (1,1) and (2,2)
inversion transition lines of ammonia
\citep*{KWT}. The ammonia survey has revealed 10 dense cores, occupying
some 7\,\% of the mapped area. Their total mass is $\sim$80\,M$_{\sun}$,
about 6\,\% of the mass traced by C$^{18}$O.

\citet*{KAY} performed a near infrared imaging survey and search for Herbig--Haro
objects in L\,1340. They established that RNO\,7 is a small embedded cluster
consisting of some 26 members, and identified three Herbig--Haro flows in
L\,1340\,A. The source of HH\,487 is probably IRAS 02224+7227, coinciding in
position with an M type  weak-line T~Tauri star \citep{Kun06}.
HH\,488 originates from RNO\,7, and HH\,489
from the Class~I source IRAS~02250+7230.

\begin{figure*}[ht!]
\centerline{
\includegraphics[width=12cm]{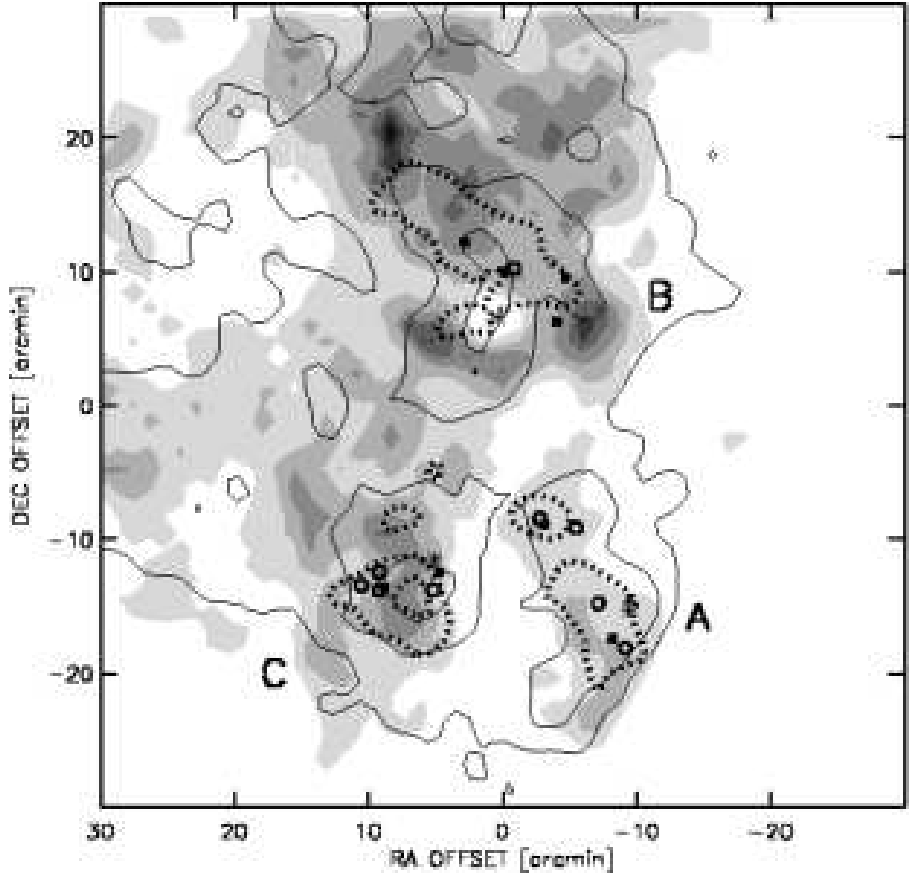}
}
\caption{$^{13}$CO (solid contours) and C$^{18}$O  (dotted contours) integrated
intensity  overlaid on the optical
extinction map (shading) of L\,1340 constructed from star counts on the DSS.
Coordinate offsets are given in arcmin with respect to
RA(2000)=2$^{\rm h}29^{\rm m}42^{\rm s}$ and
Dec(2000)=+72$^{\rm o}43^{\prime}22^{\prime\prime}$. The lowest
contour of $^{13}$CO is at 1.0\,K\,kms$^{-1}$, and the increment is 1.5\,K\,kms$^{-1}$.
The C$^{18}$O contours displayed are 0.45 and 0.75\,K\,kms$^{-1}$. Both the
lightest shade and the increment is 1\,mag. The $A_\mathrm{V}$ values
displayed are corrected for the foreground extinction. Open circles indicate
the ammonia cores, which probably represent the regions of highest volume densities.
Dots are optically invisible IRAS point sources, and asterisks show the
positions of the RNOs.}
\label{Fig_1340}
\end{figure*}

\begin{figure*}[ht!]
\centerline{
\includegraphics[width=12cm]{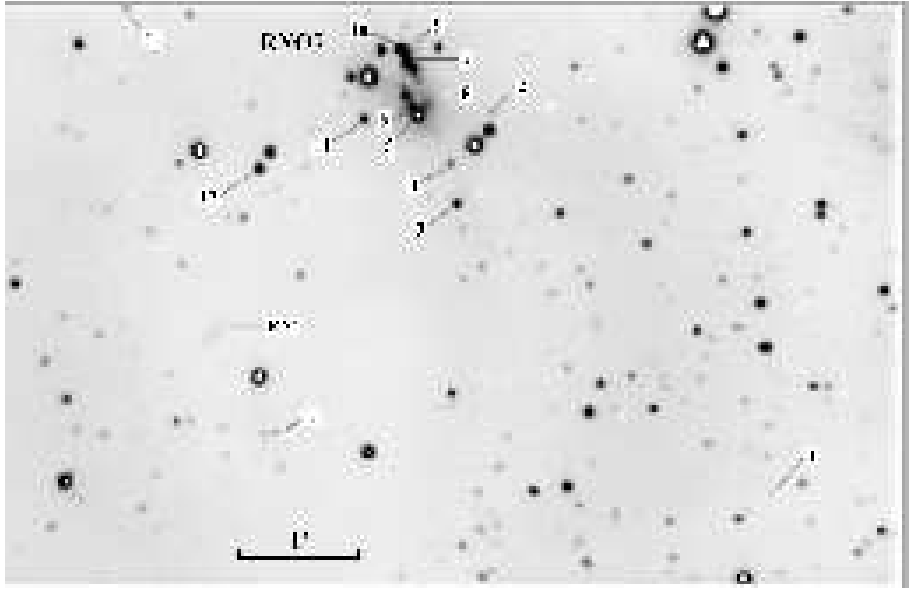}
}
\caption{H$\alpha$ emission stars and a reflection nebula (RN\,1) detected by \citet{MMN03}
near RNO\,7.}
\label{Fig_MMN}
\end{figure*}

\citet{MMN03} identified further HH objects, HH\,671 and HH\,672, originating
from RNO\,7. They detected 14 H$\alpha$ emission stars in the region of RNO\,7.
Table~\ref{Tab_1340_star} lists the IRAS point sources and H$\alpha$ emission
stars selected as candidate YSOs by \citet{KOS94} and \citet{MMN03}, and
Table~\ref{Tab_HH} lists the known HH objects and sources in L\,1340.
Figure~\ref{Fig_1340} shows the surface distribution
of $^{13}$CO, C$^{18}$O, and visual extinction, and  positions of ammonia
cores, IRAS point sources, and RNOs in L\,1340, adopted from \citet{KWT}.
Figure~\ref{Fig_MMN}, adopted from \citet{MMN03}, is the finding chart
of the H$\alpha$ emission stars near RNO\,7.

\citet{OLinger} mapped  L\,1340\,B at 450 and 800\micron \ using SCUBA on JCMT.
They discovered 25 submillimeter sources. By mapping the region in the J=3-2
transition of CO, they discovered outflows around several sources.
The submm sources  are listed in Table~\ref{Tab_L1340_smm}. There are several
Class~0 protostars among them.  \citet{OLinger}
conclude that these sources represent a second generation of stars born in the
cloud, and their formation might have been triggered by the  Herbig Be star
KOS94 R3b, located in the same cloud.

\begin{table}[h!]
\caption{Herbig-Haro objects in L\,1340}
\label{Tab_HH}
\smallskip
\begin{center}
{\footnotesize
\begin{tabular}{lcclll}
\tableline
\noalign{\smallskip}
Object 	 & RA(J2000) & Dec(J2000) & Comments & Source & Ref. \\
\noalign{\smallskip}
\tableline
\noalign{\smallskip}
HH 487A	 &   02 26 19.0 & 72 34 42 & Bow shocks & IRAS 02224+7227 & 1 \\
HH 488 S1 &  02 28 00.3 & 72 35 58 & Source plus jet & & 1 \\
HH 488 S2 &  02 28 00.0 & 72 35 58 & Source plus jet & & 1 \\
HH 671A &    02 28 16.1 & 72 37 45 & && 2 \\
HH 671B  &   02 28 09.2 & 72 36 28 &&& 2 \\
HH 488A	&    02 28 22.5 & 72 34 56 & Flow 1 & & 1,2 \\
HH 488B  &   02 28 30.6 & 72 34 37 & & & 2 \\
HH 488C	&    02 28 40.0 & 72 34 16 & Flow 1 & & 1 \\
HH 488D	&    02 28 44.0 & 72 35 37 & Flow 2 & &  1 \\
HH 488E &    02 28 38.6 & 72 34 30 & & & 2 \\
HH 489A	&    02 29 37.7 & 72 44 50 & North knot & IRAS 02250+7230 & 1 \\
HH 672   &   02 28 53.2 & 72 36 13 & & & 2 \\
\noalign{\smallskip}
\tableline
\end{tabular}}
\end{center}
\hskip5mm{\footnotesize {\it References\/:} 1 -- \citet{KAY}; 2 -- \citet{MMN03}.}
\end{table}

\begin{table}
\caption{IRAS sources and H$\alpha$ emission stars in L\,1340}
\label{Tab_1340_star}
\smallskip
\begin{center}
{\footnotesize
\begin{tabular}{l@{\hskip1mm}c@{\hskip1mm}r@{\hskip1mm}r@{\hskip1mm}r@{\hskip2mm}l}
\tableline
\noalign{\smallskip}
Names & 2MASS/RA,DEC & J & H & K$_s$ & Type$^{*}$ \\
\noalign{\smallskip}
\tableline
\noalign{\smallskip}
\multicolumn{6}{c}{L\,1340\,A} \\
\noalign{\smallskip}
\tableline
\noalign{\smallskip}
IRAS 02224+7227, HH\,487\,S  &   02270555+7241167  & 9.901 &  9.084 &   8.579 & WTTS \\
MMN03 H$\alpha$\,1  & 02273240+7234040  &      \\
HH\,488 S2 & 02275976+7235561 & 16.763 & 16.101 & 15.348 & \\
HH\,488 S1 & 02280134+7236104 & 16.202 & 15.372 & 15.199 \\

MMN03 H$\alpha$\,2                &        02280700+7237345 & 14.873 & 13.415 &  12.600  \\
MMN03 H$\alpha$\,3               &         02281182+7236447 & 13.110 & 11.695 & 10.784  \\
MMN03 H$\alpha$\,4               &         02281259+7237067 & 15.158 & 14.027&  13.536 \\
IRAS 02236+7224,&  02281661+7237328 & 11.715 & 10.406 &  9.247  \\
~~KOS94 H$\alpha$\,1, MMN03 H$\alpha$\,5 \\
MMN03 H$\alpha$\,6               &         02281748+7237384 & 14.070 & 12.893 & 12.278 \\
MMN03 H$\alpha$\,7               &         02281782+7238009 & 12.298 & 10.977 &  9.903 \\
MMN03 H$\alpha$\,8               &         02281840+7237479 & 15.077 & 13.878 & 13.196 \\
KOS94 H$\alpha$\,2b, MMN03 H$\alpha$\,9 &  02281877+7238091 & 12.397: &  12.223 & 11.715 \\
KOS94 H$\alpha$\,2, MMN03 H$\alpha$\,10 &  02281818+7238069 & 12.194:&  11.385 & 10.451 & CTTS \\
IRAS 02238+7222  &  02283280+7235270  &     \\
MMN03 H$\alpha$\,11              &         02282357+7237317 & 13.596 & 12.499 & 12.020 \\
MMN03 H$\alpha$\,12              &         02283719+7237061 & 13.085 &  12.197&  11.820 \\
MMN03 H$\alpha$\,13 &  02283760+7234430 &     \\
KOS94 H$\alpha$\,5       &                 02285180+7239143 & 12.933  & 11.989 & 11.326 & CTTS \\
MMN03 H$\alpha$\,14                &       02285420+7238352 & 14.538 &  13.493 & 12.834 \\
IRAS 02250+7230, HH 489\,S  &    02294309+7243597 & 16.028 & 14.898  & 13.914 \\
\noalign{\smallskip}
\tableline
\noalign{\smallskip}
\multicolumn{6}{c}{L\,1340\,B} \\
\noalign{\smallskip}
\tableline
\noalign{\smallskip}
IRAS 02232+7250, L1340\,B\,smm\,1  &   02280003+7304073 & 14.463 & 13.101 & 12.542 \\
IRAS 02232+7250, L1340\,B\,smm\,1  &   02280074+7304154 & 15.493 &  13.476 & 11.895 \\
KOS94 R3b  &  02292062+7304514 & 11.771 &  11.702  & 11.576 &  HBe \\
KOS94 H$\alpha$\,9, L1340B\,smm\,8  & 02292109+7258120 & 12.160 & 11.428 &  11.003 & CTTS \\
IRAS F02246+7248, L1340B\,smm\,10 &  02292330+7302220  &    \\
IRAS 02247+7245, L1340B\,smm\,11   &  02292910+7259040 &  \\
KOS94 H$\alpha$\,4   &   02293037+7311429 & 12.821 & 11.523 &  10.658 & CTTS \\
IRAS 02256+7249  &  02302610+7302460  &     \\
IRAS 02259+7246,  RNO\,8   &    02303247+7259177 & 13.628 &  12.229  & 10.899 & CTTS  \\
RNO\,8 s1      &    02303681+7259566 & 13.309 & 12.511 & 12.134 & CTTS \\
RNO\,8 s2      &    02303911+7259572 & 12.026 & 10.537 &  9.403 & CTTS \\
IRAS 02263+7251  &  02310640+7304580  &  \\
\noalign{\smallskip}
\tableline
\noalign{\smallskip}
\multicolumn{6}{c}{L\,1340\,C} \\
\noalign{\smallskip}
\tableline
\noalign{\smallskip}
IRAS 02267+7226           &   02312688+7240191 & 15.114: & 13.757 & 12.699 \\
IRAS 02267+7226            &  02312871+7240158 & 18.501: & 16.190 & 14.765 \\
RNO\,9 s2        &    02313994+7241575 & 13.826 & 12.658 &  12.098 & CTTS \\
RNO\,9           &          02314031+7241419 & 11.499 & 10.582  &  9.831 & F3IVe  \\
IRAS 02276+7225  &  02322210+7239020  &    \\
IRAS F02277+7226 & 02322888+7240561 & 15.589 & 14.948 & 14.726 \\
IRAS F02279+7225, KOS94 H$\alpha$\,10  &     02323897+7239038 & 12.291 & 11.081 &  10.209 \\
KOS94 H$\alpha$\,11   &    02330153+7243269 & 16.494 & 14.675 &  12.921 \\
KOS94 H$\alpha$\,13  &     02350799+7251034 & 12.936&  11.950 &  11.218 & CTTS \\
\noalign{\smallskip}
\tableline
\noalign{\smallskip}
\multicolumn{4}{l}{\parbox{0.8\textwidth}{\footnotesize
    KOS94\,H$\alpha$: H$\alpha$ emission stars listed in \citet{KOS94};}}\\[1ex]
\multicolumn{4}{l}{\parbox{0.8\textwidth}{\footnotesize
    MMN03 H$\alpha$:  H$\alpha$ emission stars listed in \citet{MMN03}.}}\\[1ex]
\multicolumn{4}{l}{\parbox{0.8\textwidth}{\footnotesize
    $^{*}$Kun et~al., in preparation.}}
\end{tabular}}
\end{center}
\smallskip
\end{table}

\begin{table}[!ht]
\caption{Submillimeter sources in L\,1340\,B  from \citet{OLinger}}
\label{Tab_L1340_smm}
\smallskip
\begin{center}
{\footnotesize
\begin{tabular}{l@{\hskip2mm}c@{\hskip1mm}c@{\hskip2mm}l}
\tableline
\noalign{\smallskip}
Name & RA(J2000) & Dec.(J2000)  & remarks \\
\noalign{\smallskip}
\tableline
\noalign{\smallskip}
L1340\,B\,smm\,1 & 02 27 55.3 & 73 04 15 & IRAS F02232+7250?, multiple (5 peaks) \\
L1340\,B\,smm\,2 &  02 28 07.1 & 72 59 08 &  multiple (4 peaks) \\
L1340\,B\,smm\,3 & 02 28 08.9 &	73 04 07 &  multiple (3 peaks) \\
L1340\,B\,smm\,4 & 02 28 14.3 &	73 05 29 &  multiple (2 peaks) \\
L1340\,B\,smm\,5 & 02 28 19.7 &	73 05 11 &  single	 \\
L1340\,B\,smm\,6 & 02 28 52.6 & 73 01 17 &  multiple (3 peaks) no IRAS, \\
&&& no outflow, prestellar \\
L1340\,B\,smm\,7 & 02 29 05.4 & 73 01 12 &  single, no outflow, prestellar \\
L1340\,B\,smm\,8 & 02 29 18.4 & 72 58 12 &  single, no outflow,  2MASS \\
L1340\,B\,smm\,9 & 02 29 20.3 & 73 01 21 & probably outflow, no IRAS  \\
L1340\,B\,smm\,10& 02 29 20.8 & 73 02 23 & F02246+7248, 02291961+7302237, 3 peaks, \\
&&& Class 0, Class I \\
L1340\,B\,smm\,11 & 02 29 32.0 & 72 59 17 &  02248+7245, prob. Class 0 \\
L1340\,B\,smm\,12 & 02 29 56.4 & 73 02 20 &  F02252+7248, outflow, bright,  Class 0 \\
L1340\,B\,smm\,13 & 02 29 59.5 & 73 02 53 &  outflow, no IRAS, no 2MASS, Class 0 \\
L1340\,B\,smm\,14 & 02 30 04.3 & 73 02 50 &  no IRAS,  no 2MASS, prob. Class 0 \\
L1340\,B\,smm\,15 & 02 30 08.4 & 73 02 53 &  no IRAS,  no 2MASS, prob. Class 0 \\
L1340\,B\,smm\,16 & 02 30 18.7 & 73 02 48 & 3 peaks, prob. Class 0 \\
L1340\,B\,smm\,17 & 02 30 22.7 & 73 05 03 &  no IRAS,  no 2MASS, Class 0 or prestellar \\
L1340\,B\,smm\,18 & 02 30 34.9 & 73 00 17 & 3 peaks, cluster of at least 6 2MASS sources \\
L1340\,B\,smm\,19 & 02 30 35.3 & 73 03 42 &  at least 3 peaks, outflow, Class 0+Class I \\
L1340\,B\,smm\,20 & 02 30 41.6 & 73 01 59 & no IRAS,  no 2MASS, prob. Class 0 \\
L1340\,B\,smm\,21 & 02 30 42.3 & 73 03 10 &  outflow, faint 2MASS, prob. Class 0 \\
L1340\,B\,smm\,22 & 02 30 43.8 & 73 04 24 &   prob. Class 0 or Class I \\
L1340\,B\,smm\,23 & 02 30 45.0 & 73 01 39 &   outflow,  no IRAS, no 2MASS, Class 0 \\
L1340\,B\,smm\,24 & 02 30 55.1 & 72 58 07 & no IRAS, no outflow \\
L1340\,B\,smm\,25 & 02 31 02.3 & 72 58 09 & outflow, no IRAS, no 2MASS, Class 0 \\
\noalign{\smallskip}
\tableline
\end{tabular}}
\end{center}
\end{table}

\section{Star Forming Regions at $\mathbf{|b| < 10\deg}$}

The pre-main sequence stars in Cassiopeia at $|b| < 10\deg$ are listed
in Table~\ref{Tab_pms}, and the outflow-driving objects are shown in
Table~\ref{Tab_outflow}.

\begin{table}[ht!]
\caption{Pre-main sequence stars in Cassiopeia located at $|b| < 10\deg$}
\label{Tab_pms}
\smallskip
\begin{center}
{\footnotesize
\begin{tabular}{l@{\hskip2mm}cc@{\hskip2mm}c@{\hskip2mm}l@{\hskip2mm}l@{\hskip2mm}l}
\tableline
\noalign{\smallskip}
Names & IRAS & RA(J2000) & Dec(J2000) & Type  & Cloud & Ref. \\
\noalign{\smallskip}
\tableline
\noalign{\smallskip}
HBC\,317, MWC 1080 & 23152+6034 & 23 17 25.6 & 60 50 43 & Cont.& L\,1238 &  1 \\
HBC\,2, LkH$\alpha$ 197 &   & 00 10 36.4 & 58 50 05 & K7 &  L\,1265 & 1 \\
HBC\,3, LkH$\alpha$ 198 & 00087+5833 & 00 11 26.0 & 58 49 29 & A &  L\,1265 & 1 \\
HBC\,325, V376 Cas & & 00 11 26.1 & 58 50 03 & B5e & L\,1265 & 1 \\
HBC\,329, VX Cas & 00286+6142 & 00 31 30.7  & 61 58 51 & A0e & & 1 \\
RNO 1B, V710 Cas & & 00 36 46.3 & 63 28 54 & F8IIe & L\,1287 & 5,6\\
RNO 1C & &  00 36 46.8 & 63 28 58 & FU Ori & L\,1287 & 6 \\
HBC\,6, LkH$\alpha$\,200, V828 Cas & & 00 42 29.3 & 61 55 46 & K1 & L\,1291 & 1 \\
HBC\,330, BD\,+61\deg154, & 00403+6138 & 00 43 18.3 & 61 54 40 & B8 & L\,1302 & 1\\
~~MWC\,419, V594 Cas, VDB\,4 \\
HBC\,7, LkH$\alpha$\,201 && 00 43 25.3 & 61 38 23 & B3e & L\,1302 & 1 \\
HBC\,331, LkH$\alpha$ 203 && 00 44 27.4 & 62 10 46 & K5 & L\,1302 & 1  \\
HBC\,332, LkH$\alpha$ 204 && 00 45 09.9 & 62 04 26 & K8 & L\,1302 & 1 \\
GLMP\,9  & 00422+6131 & 00 45 09.9 & 61 47 57 & T Tau  & L\,1302 & 4 \\
HBC\,333, LkH$\alpha$ 205 && 00 45 26.3 & 61 38 53 & Cont.& L\,1302 & 1 \\
GLMP\,11 & 00470+6130 & 00 50 04.0 & 61 47 05 & T Tau & L\,1302 & 4 \\
S\,187\,H$\alpha$ & & 01 20 15.2 & 61 33 08 & Be & L\,1317 & 2 \\
SCP NIRS 1 & & 01 23 18.2 &  61 47 40 & T Tau & L\,1317 & 3 \\
\noalign{\smallskip}
\tableline
\noalign{\smallskip}
\multicolumn{7}{l}{\parbox{0.9\textwidth}{\footnotesize
    {\it References\/:} 1 -- \citet{HBC}; 2 -- \citet{Zavagno};
    3 -- \citet{Salas}; 4 -- \citet{GLMP};  5 -- \citet{Staude};
    6 -- \citet{Kenyon}.
}}
\end{tabular}}
\end{center}
\end{table}

\begin{table}[h!]
\caption{Outflow driving sources in Cassiopeia at $|b| < 10\deg$}
\label{Tab_outflow}
\smallskip
\begin{center}
{\footnotesize
\begin{tabular}{llccl}
\tableline
\noalign{\smallskip}
Object & IRAS & RA(J2000) & Dec(J2000) &  Ref. \\
\noalign{\smallskip}
\tableline
\noalign{\smallskip}
MWC\,1080 & 23152+6034 & 23 17 25.6 & 60 50 43 & 1,2,3 \\
LkH$\alpha$~198 & 00087+5833 & 00 11 26.0 & 58 49 29 & 1,2 \\
M\,120.1+3.0 & 00213+6530 & 00 24 10.4 & 65 47 01 & 9 \\
M\,120.1+3.0 & 00259+6510 & 00 28 50.9 & 65 26 46 & 9 \\
L\,1287 & 00338+6312 & 00 36 47.5 & 63 29 02 & 4,5 \\
L\,1293 & 00379+6248 &  00 40 53.5 & 63 04 53 & 6 \\
S\,187\,IRS & & 01 23 15.0 & 61 48 47 &  7 \\
SCP NIRS 1 & & 01 23 18.2 &  61 47 40 &  8 \\
\noalign{\smallskip}
\tableline
\noalign{\smallskip}
\multicolumn{5}{l}{\parbox{0.6\textwidth}{\footnotesize
{\it References\/:} 1 -- \citet{Canto}; 2 -- \citet{Levreault}; 3 -- \citet{Koo};
4 -- \citet{Yang91}; 5 -- \citet{Snell}; 6 -- \citet{Yang90}; 7 -- \citet{Bally83};
8 -- \citet{Salas}; 9 -- \citet{Yangetal90}.}}
\end{tabular}}
\end{center}
\end{table}

\subsection{LkH$\mathbf{\alpha}$ 198 and its Environment in Lynds 1265}

\begin{figure}[htbp]
\centerline{
\includegraphics[width=6cm]{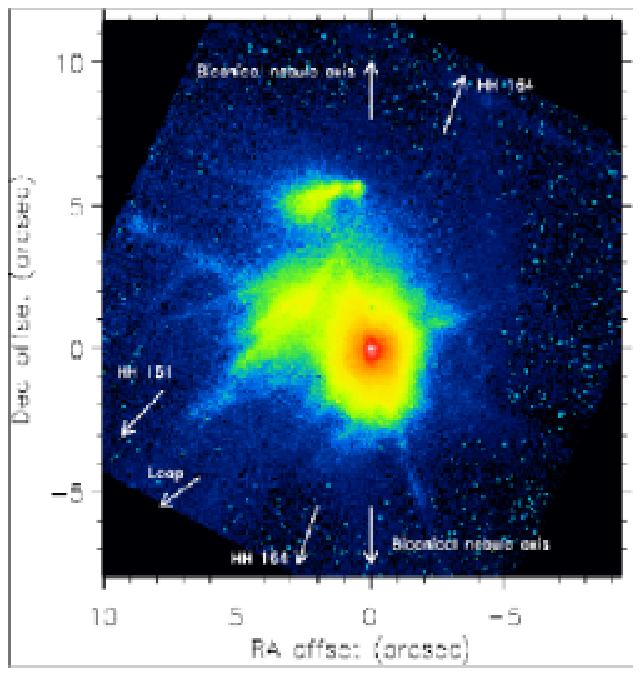}
\includegraphics[width=6cm]{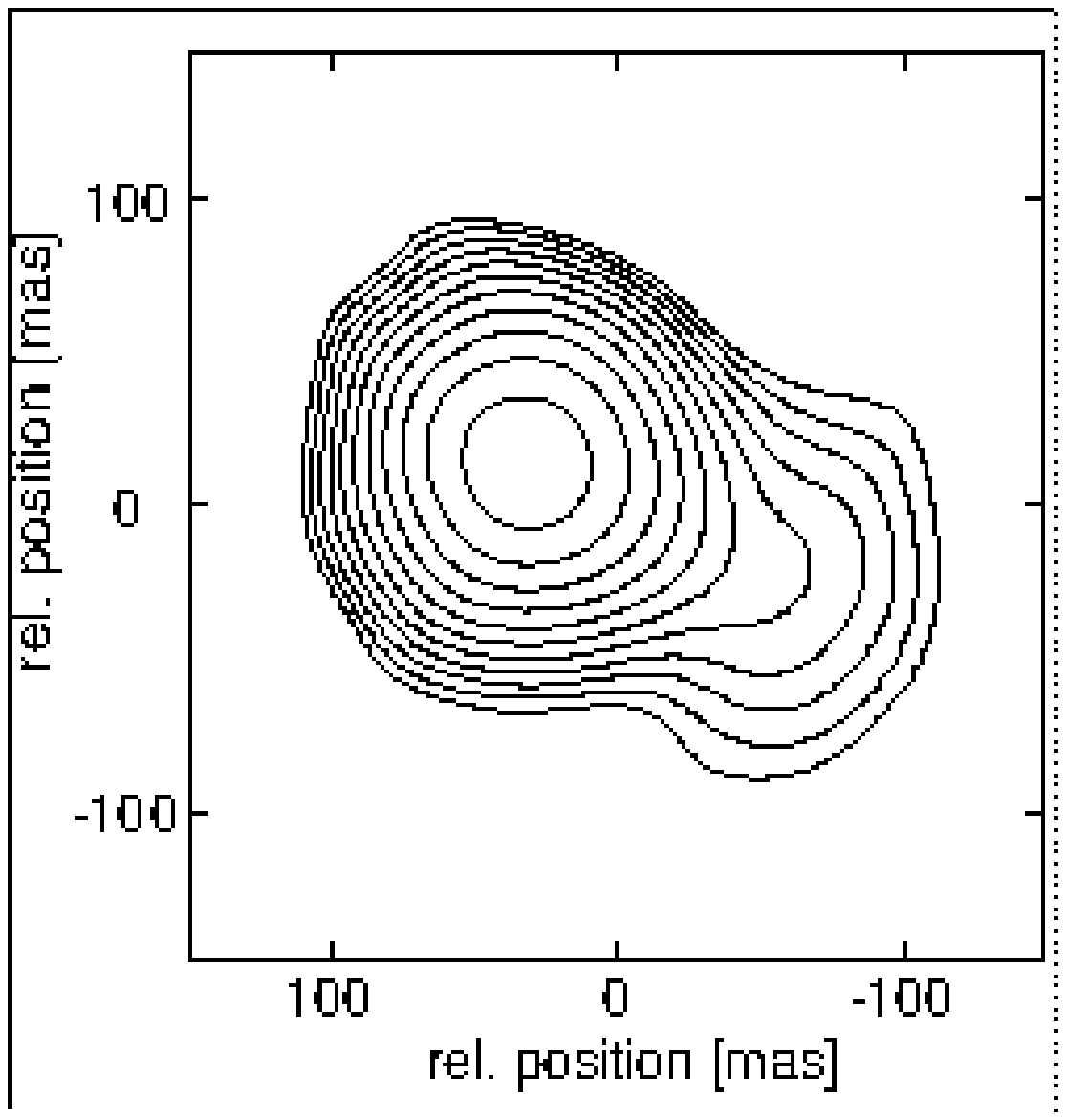}
}
\caption{{\it Left: } HST image of LkH$\alpha$\,198, taken with the NICMOS camera at 1.1\,\micron.
The source approximately 6\,\arcsec \ due north of LkH$\alpha$\,198 is the
companion LkH$\alpha$~198\,B. A fan-shaped trail of emission is visible to the east
of this companion. The speckle companion, LkH$\alpha$\,198~A2, lies within
60~mas of the primary and is not discernible in this image. The directions to various
associated HH objects and other structures are indicated. {\it Right\/}: contour map of the
LkH$\alpha$\,198 system constructed from speckle observations shows LkH$\alpha$\,198~A2
(adopted from \citet{Smith05}).}
\label{Fig_Lkha198}
\vspace{2ex}
\centerline{
\includegraphics[width=12cm]{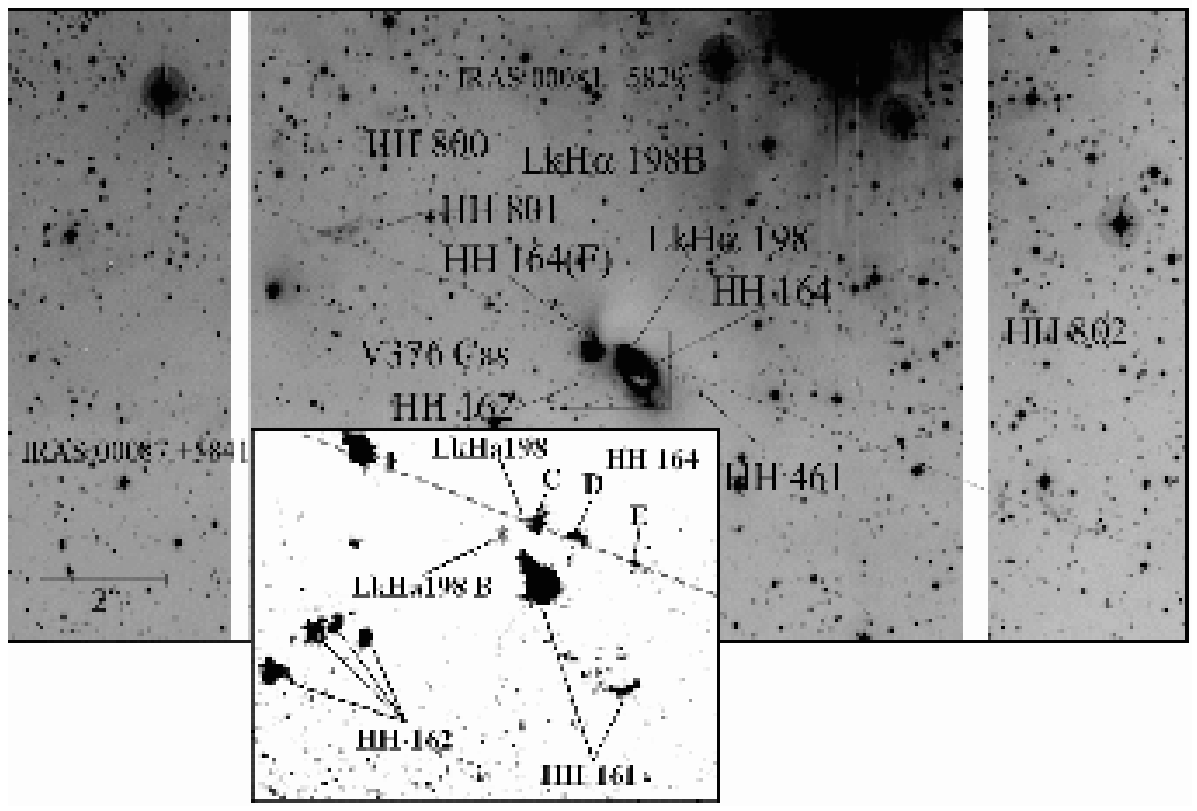}
}
\caption{LkH$\alpha$198 [SII]: Mosaic of the entire outflow around LkH$\alpha$\,198,
adopted from \citet{McGroa04}. North is to the left and west to the top in this figure.}
\label{Fig_Lkha198_hh}
\end{figure}

The  pre-main sequence star LkH$\alpha$ 198 (V633 Cas, HBC~3),
associated with the dark cloud L\,1265, was first identified by \citet{Herbig}.
\citet{CK79} classified its spectral type as B3, \citet{Hillen92} as A5, and
\citet{Hernan04} as B9. \citet{Chav85} discovered the photometric variability,
in particular, flare activity of  LkH$\alpha$\,198. He derived a distance of 600\,pc
and a luminosity of 160\,$L_{\sun}$  for LkH$\alpha$\,198, and established
that the star is some 2.5\,mag above the main sequence.
Its eventual spectral type on the main sequence is
expected to be about B3. V376~Cas, another Herbig Be star, is located 37\arcsec \
north of LkH$\alpha$~198 within the same cloud.
\citet{Lagage} imaged the region around LkH$\alpha$\,198 at 10\,$\mu$m, and
discovered a deeply embedded companion 6\arcsec \ north of the star, which they suggest
to be responsible for most of the far-infrared emission in the region.
The mid-infrared companion was also detected in the near-infrared by
\citet{Li94} and \citet{Leinert97}. The projected separation of the binary is 3300~AU.
The estimated luminosity of the companion is 100\,L$_{\sun}$, thus it could be a
third Herbig~Ae/Be star in the system.

\citet{Canto} and \citet{Levreault} found a large CO outflow from the system.
\citet{Strom} discovered an HH object, HH\,161, south-east of LkH$\alpha$\,198.
\citet{Corcoran} found two further HH flows, HH\,162, associated with V376~Cas, and
HH\,164. They identified LkH$\alpha$\,198 as the driving source of HH\,164, and the
embedded companion, LkH$\alpha$\,198\,B, as the source of HH\,161.
\citet{AR2000}  by optical imaging discovered  a bright bow shock, HH\,461, along the
axis defined by the previously known HH~164 jet.

\citet{SW94} presented submillimeter and millimeter line and continuum observations
of the region around  LkH$\alpha$~198 and V376~Cas. The continuum data revealed the
presence of a very cool object, LkH$\alpha$~198~MM, located 19\arcsec \ northwest
of LkH$\alpha$~198. LkH$\alpha$~198~MM is not visible in the near-infrared. The CO maps
suggest that the mm-source, rather than LkH$\alpha$~198 or its infrared companion
LkH$\alpha$~198\,B, may drive the large CO outflow seen in this cloud,
or that both the stars and the millimeter source drive outflows which are roughly parallel
to each other. In addition V376~Cas appears to power a small bipolar CO outflow.

\citet{BN95} modeled the extended far-infrared emission, measured by KAO, from the
system, and established that the source of the emission is the optical star LkH$\alpha$\,198,
and not its infrared companion. High resolution interferometry toward LkH$\alpha$\,198 and
the surrounding region by \citet{diFran97} at 2.7~mm failed to detect LkH$\alpha$\,198,
but LkH$\alpha$\,198~MM was detected and resolved.
\citet{Hajjar00} revisited the probable protostellar object LkH$\alpha$ 198~MM
northwest of LkH$\alpha$~198.
They found it to be the main source of emission in the region at millimeter wavelengths
and probably the major contributor to the submillimeter radiation down to 100\,$\mu$m,
and the most likely source of the large-scale CO outflow.

The stars are embedded in a reflection nebulosity (see the left panel of
Fig.~\ref{Fig_Lkha198}), which was also detected in the near infrared
by \citet{Testi98}. The NIR nebulosity prevented \citeauthor{Testi98} from
detecting a possible group of low-mass stars, usually present around
Herbig~Be stars.

\citet{Fukagawa02} carried out high-resolution near-infrared imaging
of LkH$\alpha$ 198
using the Coronagraphic Imager with Adaptive Optics  (CIAO) on the Subaru 8.2-m Telescope.
The images resolve the infrared companion LkH$\alpha$~198\,B and an associated parabolic
reflection nebula, indicating that it is an outflow source. LkH$\alpha$~198\,B exhibits
near-infrared colors suggestive of a Class~I-like source.
Faint nebulae associated with outflow activity and a bright flattened envelope were
also found around the primary.

\citet{McGroa04} discovered  a 2~pc long optical outflow powered by
LkH$\alpha$ 198. They
discovered new Herbig-Haro objects, HH\,800, HH\,801, HH\,802, originating from the system.
Fig.~\ref{Fig_Lkha198_hh}, adopted from \citet{McGroa04} shows all
the HH objects around the LkH$\alpha$~198 system.
\citet{Perrin04} have used laser guide star adaptive optics and a near-infrared dual-channel
imaging polarimeter to observe light scattered in the circumstellar environment of
Herbig Ae/Be stars on scales of 100 to 300 astronomical units. They detected a strongly
polarized, biconical nebula 10~arcsec (6000~AU) in diameter around
LkH$\alpha$~198 and also observed a polarized jet-like feature associated with
the deeply embedded source LkH$\alpha$~198-IR.
\citet{Smith05} detected a close optical companion of LkH$\alpha$~198,
based on diffraction-limited bispectrum speckle interferometry observations
(right panel of Fig.~\ref{Fig_Lkha198}). The new object, LkH$\alpha$~198~A2 lies
at a separation of approximately 60~mas, or 36~AU from the brighter component. The plane
of its orbit appears to be significantly inclined to the plane of the
circumprimary disk, as inferred from the orientation of the outflow.

\citet{Matthews07} combined $^{12}$CO  data from FCRAO with high resolution BIMA array data to
achieve a naturally weighted synthesized beam of $6.75\arcsec \times 5.5\arcsec$
toward LkH$\alpha$\,198. They found that the outflow around LkH$\alpha$\,198 resolves into
at least four outflows, none of which are centered on LkH1$\alpha$\,98--IR.

\subsection{L\,1287}

\begin{figure}[bp]
\centerline{
\includegraphics[width=8cm]{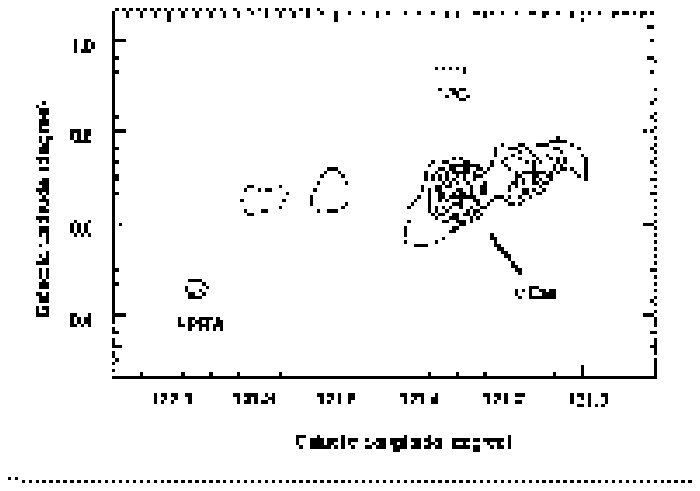}
\includegraphics[width=5cm]{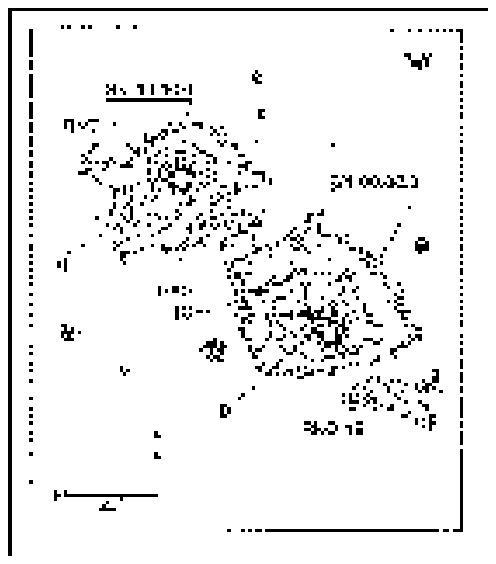}
}
\caption{{\it Left\/:} $^{13}$CO integrated intensity map of L\,1287, observed by \citet{Yang91}. Contours start at
1\,K\,km\,s$^{-1}$ with steps of 1\,K\,km\,s$^{-1}$. IRAS 00338+6312 is
denoted by cross near the center of the main peak. The cross at the west
peak represents another source, IRAS 00332+6315. The direction of $\kappa$ Cas, the
assumed source of triggered formation of the dense core,
is illustrated by the arrow. {\it Right\/:}  I-band image of the RNO~1 region
\citep{Staude}}
\label{Fig_L1287}
\end{figure}

L\,1287 (TDS\,463) is a filamentary dark cloud, located at 850\,pc from the Sun \citep{Yang91},
and extending over 10~pc along the Galactic plane. \citet{Yang91} mapped the
cloud in  $^{13}$CO, HCO$^{+}$, and HCN lines. The total mass of the cloud was estimated
to be some 240\,M$_{\sun}$ from the $^{13}$CO observations. It contains at least four separate
cores, aligned along the filament. The region of strongest CO emission,
{\em L\,1287 main core\/}, contains a mass of some 13\,M$_{\sun}$, according to
the $^{13}$CO observations. The IRAS point source IRAS~00338+6312 is projected
on the peak CO position of L\,1287  \citep[Fig.~\ref{Fig_L1287} left panel, adopted from][]{Yang91}.
An energetic  bipolar molecular outflow has been detected around this source
\citep{Snell,Yang91}, and it is associated with  H$_2$O \citep{Henning,Fiebig95,Fiebig96},
OH \citep*{Wouterloot93}, as well as methanol \citep*{Slysh99,Kalenskii} maser emission.

\citet{Walker}  presented HCO$^{+}$ (J=1--0) and CS (J=2--1 and 5--4) data for the
molecular cloud centred on IRAS 00338+6312. They found that the line profiles suggest a
collapsing cloud with density and infall velocity increasing towards the centre
as $r^{-3/2}$ and $r^{-1/2}$, respectively, in accordance with the predictions
by \citet{Shu77}.
\citet{Mookerjea} observed IRAS 00338+6312 using the two-band FIR photometer system at the
Cassegrain focus of the TIFR 100~cm (f/8) balloon-borne telescope. An area of
 $22\arcmin \times 7\arcmin$ centered on IRAS 00338+6312 was mapped
in the FIR bands at 143\,\micron \ and 185\,\micron.  The source was resolved
at both wavelengths.

L\,1287 harbors the nebulous young star RNO\,1 \citep[F5e,][]{RNO}. \citet{Staude} discovered
a further young stellar object located at 50\arcsec\ south-west of RNO~1 (0.2\,pc at a
distance of 800~pc) which they called RNO~1B. Based on optical spectroscopic observations
they have shown that RNO~1B is a \normalsize{FU Orionis} type star.
Further faint, probably deeply embedded stars can be identified in their
$I$-band image near RNO~1 (right panel of Fig.~\ref{Fig_L1287}). Near-infrared spectroscopic
observations by \citet{Kenyon} confirmed the FUor nature of RNO~1B, and have
shown that  RNO~1C, located at 6\arcsec\ from RNO~1B, is also a FUor.
\citet{Kenyon} suggest that the binary system RNO~1B/RNO~1C is the driving source
of the molecular outflow.

IRAS 00338+6312 is 5\arcsec \ and 11\arcsec \ from  RNO\,1C and RNO\,1B, respectively.
A number of observations support that the catalog coordinates of IRAS
00338 +6312 accurately
mark the location of a deeply embedded YSO, distinct from the FUors. Using polarimetric
imaging, \citet{Weintraub93} found an embedded protostar that they suggested was the star
driving the molecular outflow, since the position of the protostar was closer than
either RNO\,1B or 1C to the position of the IRAS source. VLA observations by
\citet{Anglada} revealed four 3.6~cm sources within 30\arcsec \ of
the IRAS position, all four of which are possibly PMS objects embedded in the L\,1287
cloud core. The position of the strongest VLA source, VLA~3, coincides within the
errors with the position of the protostar found by  \citeauthor{Weintraub93}, while a second
of the VLA sources, VLA~1, appears to coincide with the position of RNO~1C.
\citet{McMuldroch} using the OVRO millimeter array detected a millimeter continuum source
at both 3.1 and 2.6~mm, whose position coincides within the errors with that of RNO\,1C.
Neither RNO\,1B nor VLA~3 were detected in the millimeter continuum, a result which would
argue for RNO\,1C being the source driving the outflow.
The 850 and 450 \micron \ SCUBA maps obtained by \citet{SW01} show a strong elliptical
dust source (RNO\,1~SM\,1) centered very close to the position of RNO\,1C. The long
axis of RNO\,1~SM\,1 ($\sim +18\deg$) is roughly orthogonal to the position angle
of the CO outflow ($\sim -50\deg$). In addition to the large central source,
the maps reveal a large region of emission surrounding
RNO\,1~SM\,1 and a ridge of emission extending to the southeast.
In the high-resolution 450 \micron \ image,  an extension from  RNO\,1~SM\,1 can be seen
in the direction of the protostellar source VLA~3, supporting that that
VLA~3 also is associated with dust emission.
Along the southeast ridge, the SCUBA maps show the presence of two additional
submillimeter sources, RNO\,1~SM\,2 and RNO\,1~SM\,3, both of which are securely
detected at 1.3 mm and 850 \micron. No optical, infrared or IRAS counterparts are
known at the positions of either RNO\,1~SM\,2 or RNO\,1~SM\,3.

\begin{figure*}[htbp]
\centerline{
\includegraphics[width=12cm]{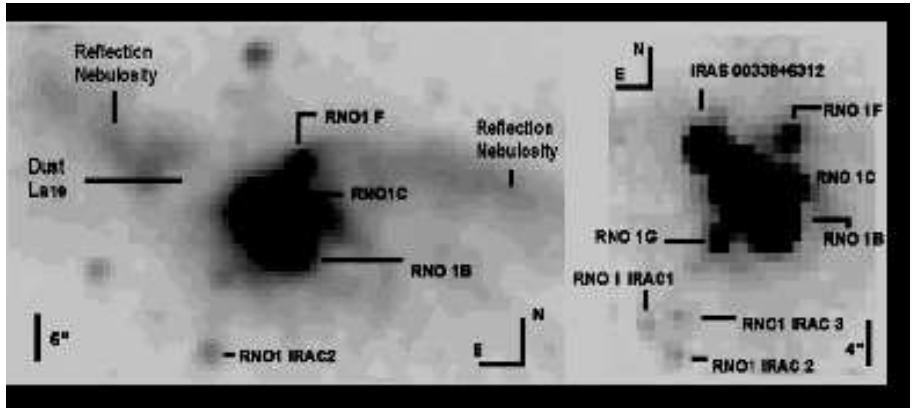}
}
\caption{{\it Left\/:} The RNO\,1B/1C region in the 2MASS K$_s$ band. {\it Right\/:} The same
region in the Spitzer IRAC  5.8\,\micron\ band. Adopted from \citet{Quanz07}.}
\label{Fig_2m_irac}
\end{figure*}

\begin{table}[htbp]
\caption{Positions of 3.8\,\micron \ sources in the L\,1287 core \citep{Weintraub96}}
\label{Tab_Wein}
\smallskip
\begin{center}
{\footnotesize
\begin{tabular}{lccc}
\tableline
\noalign{\smallskip}
Source  & Offset RA(\arcsec) &  Offset Dec(\arcsec) &  Dist.(\arcsec) \\
\noalign{\smallskip}
\tableline
\noalign{\smallskip}
\multicolumn{4}{c}{Sources Near IRAS 00338+6312}  \\
\noalign{\smallskip}
\tableline
\noalign{\smallskip}
Northern centroid  &   $-$1.50  & $-$0.50~ & 1.58 \\
VLA 3  &   $-$1.20  & 0.20  & 1.22 \\
H$_2$O maser  &   $-$1.09  & 0.10  & 1.09 \\
DLIRIM No. 4  &   $-$0.78  & 0.50  & 0.93 \\
IRAS 00338+6312  &   ~0.00  & 0.00 &  0.00 \\
DLIRIM No. 1  &   ~1.40  & 0.68 &  1.57 \\
\noalign{\smallskip}
\tableline
\noalign{\smallskip}
\multicolumn{4}{c}{Other Sources} \\
\noalign{\smallskip}
\tableline
\noalign{\smallskip}
RNO 1F  &   $-$9.60  & ~~3.09  & 10.39 \\
RNO 1B  &   $-$8.10  & ~$-$8.00  & 11.38 \\
RNO 1D  &   $-$3.92  & ~$-$6.14 &  ~7.28 \\
RNO 1C  &   $-$3.88  & ~$-$3.52 &  ~5.24 \\
VLA 4  &   $-$0.96  & ~$-$8.40  & ~8.45 \\
DLIRIM No. 3  &  $-$0.73  & $-$10.62  & 10.65 \\
Southern centroid  &   $-$0.15  & ~$-$6.00 & ~6.00 \\
DLIRIM No. 2  &  ~~0.18  & ~~5.12  & ~5.12 \\
\noalign{\smallskip}
\tableline
\end{tabular}}
\end{center}
\end{table}

\begin{table}[htbp]
\caption{Positions and apparent brightnesses of IRAC sources in the L\,1287 core \citep{Quanz07}}
\label{Tab_Spitzer}
\smallskip
\begin{center}
{\footnotesize
\begin{tabular}{l@{\hskip2mm}l@{\hskip2mm}c@{\hskip2mm}c@{\hskip1mm}c@{\hskip1mm}c@{\hskip1mm}c@{\hskip1mm}c}
\tableline
\noalign{\smallskip}
No. &  Object name & RA(J2000) & Dec(J2000) & 3.6$\,\mu$m & 4.5$\,\mu$m & 5.8$\,\mu$m & 8.0$\,\mu$m \\
& & & & [mag] &  [mag] & [mag] & [mag] \\
\noalign{\smallskip}
\tableline
\noalign{\smallskip}
1 & RNO 1B & 00 36 46.05 & +63 28 53.3  & ~7.16  & ~6.67  & ~5.76  & 5.01 \\
2 & RNO 1C & 00 36 46.65 & +63 28 57.9  & ~6.56  & ~6.04  & ~5.58  & 4.61 \\
3 & RNO 1F & 00 36 45.74 & +63 29 04.1  & 10.28  & ~9.46  & ~8.58  & 8.20 \\
4 & RNO 1G & 00 36 47.14 & +63 28 49.9  & $\cdots$ & 10.33  & ~8.74  & 8.05 \\
5 & IRAS 00338+6312 & 00 36 47.34 & +6 29 01.6 & $\cdots$  & ~9.05  & ~7.19  & 6.72 \\
6 &RNO1 IRAC1 & 00 36 48.44 & +63 28 40.0  & 13.72  & 11.82  & 10.93  & 9.65 \\
7 & RNO1 IRAC2 & 00 36 47.90 & +63 28 36.3 & 12.00  & 10.86  & 10.86  & 9.82  \\
8 & RNO1 IRAC3 & 00 36 47.85 & +63 28 41.2 & 13.07 & 11.78  & 10.74  & 9.74  \\
\noalign{\smallskip}
\tableline
\end{tabular}}
\end{center}
\end{table}

\citet{Weintraub96}
obtained diffraction limited images of the L\,1287 core at 3.8\,\micron. In additon to
RNO\,1B/1C they identified at least six other stellar sources, as well as one region of nebulosity.
Their image revealed a 3.8\,\micron \ source coincident with IRAS 00338+6312, thus confirm that
the outflow source is a deeply embedded protostellar object.
The 3.8\,\micron \ sources identified by  \citet{Weintraub96} are listed in Table~\ref{Tab_Wein}.

\citet{Testi98} included RNO\,1B in their  search for compact clusters around Herbig~Be stars.
The star is surrounded by a bright extended nebulosity.  A small group of stars
with $r \sim 0.15$ pc was detected within the nebula.
The surface density of field stars appears to increase away
from RNO\,1B, suggesting that localized extinction is present. The size of the cluster
appears to be consistent with that of the ammonia clump detected by \citet{Estalella}.
\citet{Lorenzetti} present results of far-infrared spectroscopic observations, performed
with ISO Long Wavelength Spectrograph, for RNO\,1B.

\citet{Quanz07} presented results of Spitzer IRAC and IRS observations of L\,1287.
IRAC images of L\,1287 have revealed deeply embedded objects in the vicinity of
RNO\,1B and RNO\,1C, confirming their association with a young stellar cluster.
Eight sources have been detected in at least three of the four mid-infrared bands.
RNO\,1B/RNO\,1C are the only confirmed FUors that belong to a cluster-like environment.
The IRAC images resolve the mid-infrared source associated with IRAS
00338 +6312, which is an
intermediate-mass protostar. IRS spectra of the objects reveal their icy and dusty
circumstellar environment, as well as detect H$_2$ emission lines from purely rotational transitions.
These lines arise from shocked material within the molecular outflow.
Table~\ref{Tab_Spitzer} lists the coordinates of objects revealed by the
IRAC images \citep{Quanz07}.
The 2MASS K$_s$ and IRAC 5.8\,\micron\ images of the L\,1287 main core, adopted from
\citet{Quanz07} and displayed in Figure~\ref{Fig_2m_irac} show several faint members of the
small cluster around RNO\,1B/RNO\,1C.

\subsection{L\,1293}

L\,1293 appears as a dark cloud of opacity class 4 in \citeauthor{Lynds}'
(\citeyear{Lynds}) catalog. The three IRAS
sources with non-stellar flux density distribution, projected on the cloud, are listed
in Table~\ref{Tab_l1293}. Whereas IRAS~00364+6246 is probably a foreground star, the two
other sources may be young stellar objects associated with L\,1293.

\begin{table}[tb]
\caption{IRAS point sources projected on L\,1293}
\label{Tab_l1293}
\smallskip
\begin{center}
{\footnotesize
\begin{tabular}{lrrrrrl}
\tableline
\noalign{\smallskip}
IRAS & F$_{12}$  & F$_{25}$ & F$_{60}$ & F$_{100}$ & L$_\mathrm{IRAS}^{*}$ & Associated object \\
  & (Jy) & (Jy) & (Jy) & (Jy) &  (L$_{\sun}$) \\
\noalign{\smallskip}
\tableline
\noalign{\smallskip}
00353+6249 & 0.9 & 2.2 & 6.0 & 16.2 & 19 & 2MASS J\,00381870+6306053 \\
00364+6246 & 1.5 & 0.4 & $<$0.4 & 4.6 & $<$9 & HD\,3572 (Sp.= K7\,III) \\
00376+6248 & $<$0.3 & 0.4 & 9.9 & 26.9 & $<$21 & \\
\noalign{\smallskip}
\tableline
\noalign{\smallskip}
\multicolumn{7}{l}{\parbox{0.8\textwidth}{\footnotesize
    $^*$Assuming a distance of 850\,pc.}}
\end{tabular}}
\end{center}
\end{table}

\begin{figure*}[tb]
\centerline{
\includegraphics[width=7cm]{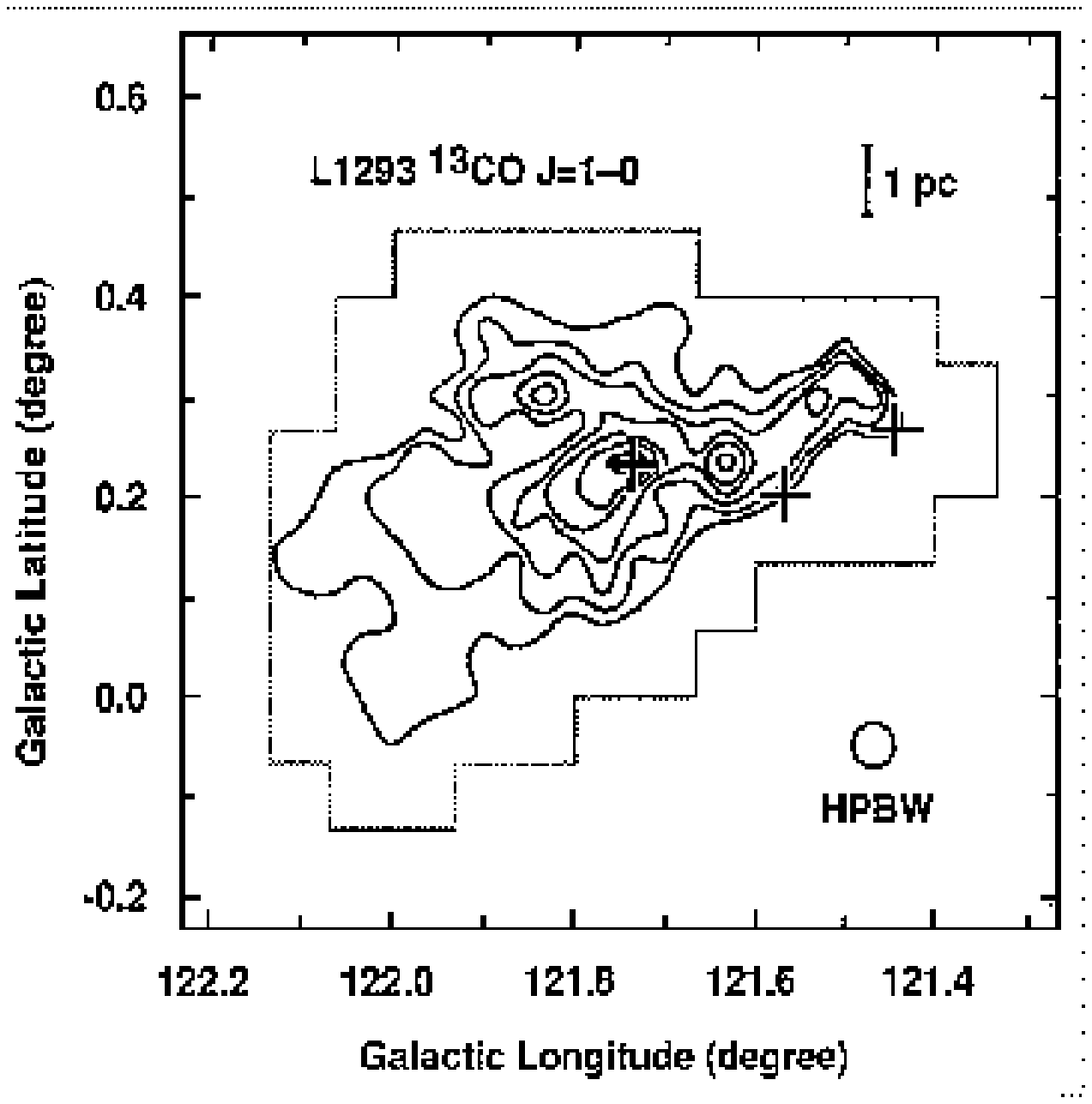}
\includegraphics[width=5cm]{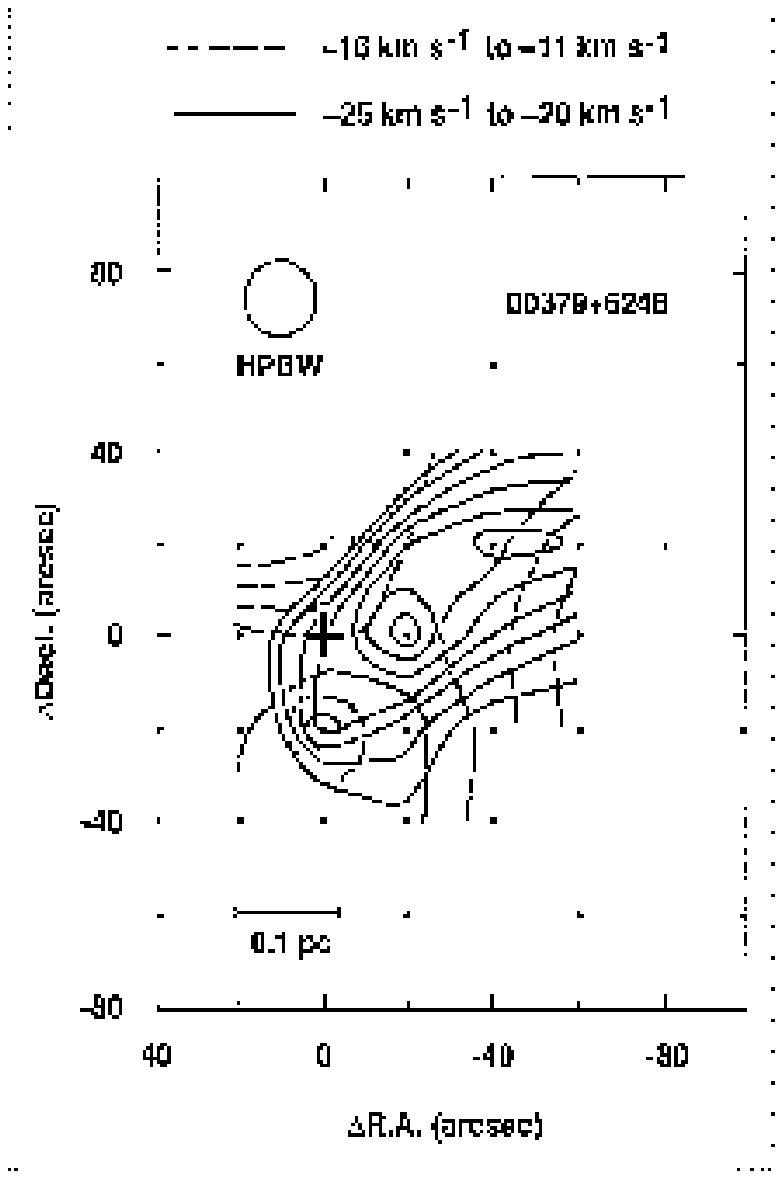}
}
\caption{{\it Left\/:} $^{13}$CO integrated intensity map of L\,1293, observed by \citet{Yang90}.
Contours start at 2\,K\,km\,s$^{-1}$ with steps of 1\,K\,km\,s$^{-1}$. The IRAS sources listed in
Table~\ref{Tab_l1293} are plotted by crosses. {\it Right\/:} A contour map of the
$^{12}$CO intensity for the high-velocity gas components. The solid lines and the grey lines
denote the blue-shifted ($-$25\,km\,s$^{-1} < v_\mathrm{LSR} < -20$\,km\,s$^{-1}$) and
the red-shifted  ($-$16\,km\,s$^{-1} < v_\mathrm{LSR} < -11$\,km\,s$^{-1}$) components, respectively.
IRAS~00376+6248 is indicated by a cross. (Adopted from \citet{Yang90}.)}
\label{Fig_l1293}
\end{figure*}

A comprehensive study of L\,1293 is presented by \citet{Yang90}. Molecular line emission
was detected at $v_\mathrm{LSR} = -18$km\,s$^{-1}$, close to those of the neighboring L\,1287 and
M\,120.1+3.0, associated with Cep~OB\,4, suggesting that these clouds are parts of the
same molecular gas stream at the same distance. Therefore \citet{Yang90} assumes a distance
of 850\,pc for L\,1293. The distribution of the $^{13}$CO emission is
shown in the left panel of Fig.~\ref{Fig_l1293}. The whole cloud, defined by the 2\,K\,km\,s$^{-1}$
contour in  $^{13}$CO, extends over $36\arcmin \times 20\arcmin$ or $9 \times 5$~pc.
The total mass of the cloud was estimated to be about 640~M$_{\sun}$. In the direction of
IRAS~00376+6248 a bipolar molecular ouflow was detected. A contour map of the $^{12}$CO intensity
for the high-velocity gas components is shown in the right panel of Fig.~\ref{Fig_l1293}.
High-resolution observations in the HCN and HCO$^{+}$ lines reveal a high-density region
around IRAS~00376+6248.

IRAS~00376+6248 appears in the CO survey by \citet{WB89} as WB\,357. It is also included
in the catalog of IRAS sources associated with CO gas in the outer Galaxy
by \citet{Kerton} as IRCO~2387. \citet{Wouterloot93} detected an H$_2$O maser emission
in the direction of IRAS~00376+6248.

\citet*{Olano} found that the space distribution and kinematics of the interstellar matter
in the region of Cep~OB4 suggest the presence of a big expanding shell, centered on
(l,b)~$\sim$~(122\deg,+10\deg). Assuming a distance of 800~pc for the center
of the shell they derived a radius of some 100~pc, expansion velocity of
4~km\,s$^{-1}$, and HI mass of $9.9 \times 10^{4}$~M$_{\sun}$ for the Cepheus~OB4~Shell.
Both L\,1287 and L\,1293 lie close to the low-latitude boundary of the Cep~OB4 Shell, plotted
in Fig.~\ref{Fig1}.

\subsection{Cas OB\,14}

L\,1287 and L\,1293 are located in a volume where the association Cas~OB\,14, defined by four
high-luminosity stars, can be found \citep{Humphreys}. Table~\ref{Tab_ob14} lists
the spectral types, Galactic coordinates and absolute magnitudes M$_V$ of these association
members. The distance of  1100~pc, given by \citet{Humphreys} for Cas~OB\,14, is the average
distance of these four stars. No lower luminosity members can be found in the literature. \citet{Snell}
and \citet{Yang91} associate L\,1287 with Cas~OB\,14, and propose that star formation in L\,1287
might have been triggered by the strong stellar winds from $\kappa$~Cas,
the most luminous member of Cas~OB\,14  (see Fig.~\ref{Fig_L1287}). As an evidence of such a trigger,
\citet{Yang91} mention that the ratio of the luminosity of the young stellar object to the mass
of the core associated with the object is much higher for the L\,1287 main core
than for isolated low-mass cores. \citet{Yang91} propose that L\,1287, L\,1293, and M\,120.1+3.0
may be parts of a giant, filamentary molecular system stretching from Cep~OB\,4 to Cas~OB\,14.

\begin{table}[htb]
\caption{Members of Cas OB14 \citep{Humphreys}}
\label{Tab_ob14}
\smallskip
\begin{center}
{\footnotesize
\begin{tabular}{llcc}
\tableline
\noalign{\smallskip}
Star &  Spectral type & {\it l / b\/} & M$_V$ \\
\noalign{\smallskip}
\tableline
\noalign{\smallskip}
HD 2905, $\kappa$ Cas & B1\,Iae & 120.8 / +0.1 & $-$7.1  \\
HD 3283 & A3\,Ib & 121.1 / $-$2.5 & $-$5.0 \\
BD\,+63\deg48 & B1\,IIIne & 120.3 / +1.7 & $-$4.5 \\
HD 2619 & B0.5\,III & 120.7 / +2.5 & $-$4.4 \\
\noalign{\smallskip}
\tableline
\end{tabular}}
\end{center}
\end{table}

\subsection{BD\,+61\deg154 and its Environment (L\,1302)}

The pre-main sequence nature of the B8-type star BD\,+61\deg154 (MWC~419, V594~Cas) was suggested by
\citet{Herbig}. The same paper reported on the discovery of seven H$\alpha$ emission
stars, namely  LkH$\alpha$~199--LkH$\alpha$~205 within 22\arcmin\ of BD\,+61\deg154.
Five of them have HBC identifications (see Table~\ref{Tab_pms}). \citet{GLMP} report
on two further T~Tauri stars in the neighborhood of BD\,+61\deg154.
\citet{Testi98} detected no density enhancement of K-band sources around
this star during their infrared search for clusterings around Herbig~Ae/Be stars.
As BD\,+61\deg154  lies on the northwestern edge of the sparse cluster NGC\,225, \citet{Herbig}
and several other authors \citep[e.g.][]{Finkenzeller,Testi98} propose that it may be
a cluster member, which would suggest a distance of 650\,pc \citep{Hagen70}. However,
according to the photometric and proper motion study by \citet{Lattanzi91}
the age of NGC~225 is about 120 million years, and BD\,+61\deg154 is not
a member of the cluster. Its Hipparcos parallax, $\pi = 3.3 \pm 1.6$ \citep{VandenA98} suggests a
smaller distance. Recently, \citet{Subra06} pointed out that 15 of the 28 members of NGC\,225,
identified by \citet{Lattanzi91} exhibit near-infrared excess, suggesting that the
cluster is probably very young and thus may contain pre-main sequence stars.

\citet{VandenA98} associate BD\,+61\deg154 with the dark cloud L\,1302.
As this region lies in the Galactic plane, several clouds along the line of sight may
contribute to the total extinction defining L\,1302. Molecular observations
confirm this hypothesis. \citet{Yang90} detected a weak CO emission from the direction of L\,1302 at
v$_\mathrm{LSR} \approx -18$\,km\,s$^{-1}$, therefore he assumed that L\,1302 belongs to the
same group of molecular clouds as L\,1287 and L\,1293. However, other
CO observations  in this region report on molecular emission at
v$_\mathrm{LSR} \approx -13$\,km\,s$^{-1}$ \citep[e.g.][]{Yonekura,Kerton}.
\citet{Kerton} detected CO emission at  v$_\mathrm{LSR} = +4.41$\,km\,s$^{-1}$
and v$_\mathrm{LSR} = -13.83$\,km\,s$^{-1}$ on the position of BD\,+61\deg154 (IRCO~2414).
\citet{Yonekura} associate their cloud $122.0-01.1$ with L\,1302 and with HBC~7 (LkH$\alpha$201).
LkH$\alpha$201 appears more distant than BD\,+61\deg154 \citep[e.g.][]{HBC,Hernan04}. Its
pre-main sequence nature is somewhat uncertain \citep{HBC,The94}.
Thus the distance of this system and the association of pre-main sequence
stars with both each other and the clouds need further studies.

\subsection{S\,187 and its Environment (L\,1317)}

S\,187 is a small optical HII region (diameter: 0\farcm9), located in the Orion arm
of the Galaxy, in the direction of the dark cloud L\,1317. Its kinematic distance is
$\sim$ 1~kpc. No other other distance information was available in the literature
until 2007, when a spectro-photometric distance of $1.44\pm0.26$~kpc was published by
\citet*{Russeil07}. They carried out spectroscopic and photometric observation
of candidate exciting stars of several HII regions, including S\,187, with the
aim to determine their distances. The result is based on spectral type and UBV
photometric data of a B2.5V type star, located at RA(2000)=$01^\mathrm{h}23^\mathrm{m}07\fs3$,
D(2000)=+61\deg51\arcmin53\farcs2 (identical with 2MASS\,J\,01230704+6151527).

S\,187 belongs to a large molecular complex, mapped by \citet*{Casoli} and
\citet*{Joncas}. A high velocity molecular outflow was discovered by \citet{Bally83}
near the core of the molecular cloud. The source of the outflow, S\,187~IRS is close
to the IRAS point source IRAS~01202+6133.

A comprehensive multi-wavelength study of S\,187 can be found in \citet{Joncas}.
They mapped the ionized, neutral and molecular components of the gas complex,
studied the infrared properties of the associated dust, identified IRAS
point sources related to star formation, and
found the possible source of the ionization by optical photometric survey of the
region. (We note that their suspected exciting source (star~4) is identical with
that used for distance determination by \citet{Russeil07}).
S\,187 is surrounded by an HI shell and a large molecular cloud.
\citet{Yonekura} derived $M=7600\,M_{\sun}$ for the
mass of the whole molecular cloud associated with S\,187.

\begin{figure*}[!ht]
\centerline{
\includegraphics[width=7cm]{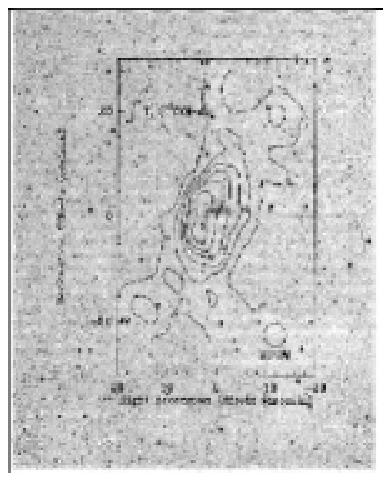}
\includegraphics[width=6cm]{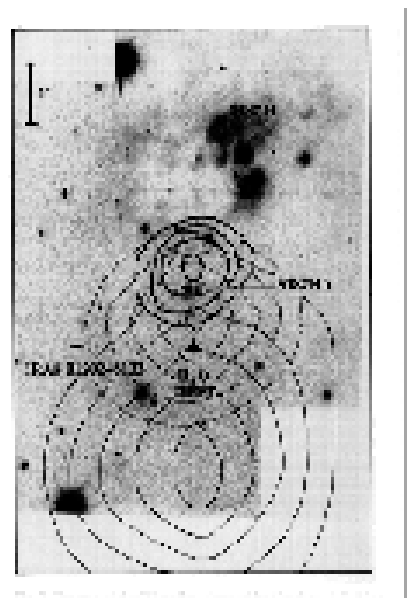}
}
\caption{{\it Left\/:}  $^{13}$CO integrated intensity map of the S\,187 molecular
cloud, superimposed on the Palomar Sky Survey image of the field  \citep{Joncas}.
The lowest contour and contour interval are 2.5\,K\,km\,s$^{-1}$. The reference position is
at 01$^\mathrm{h}19^\mathrm{m}48^\mathrm{s}$, +61\deg35\arcmin \ (1950). The nonstellar
IRAS point sources are indicated by numbers. {\it Right\/:} Map of the CO outflow observed
by \citet{Casolib} (dashed line for the
blue wing, thin lines for the red wing), superimposed on $V$ frames of the
S\,187 field obtained by \citet{Zavagno}. Ammonia emission observed by \citet{Torrelles}
is drawn by thick lines.}
\label{Fig_S187_1}
\end{figure*}

Several signposts of star formation lie in the direction of S\,187, such as
the IRAS source IRAS~01202+6133 3\arcmin \ southeast of the
HII region, and the outflow source S\,187 IRS \citep{Bally83} some 2\arcmin \
west of the IRAS source. \citet*{Henkel} discovered
H$_2$O maser emission from the direction of S\,187 IRS.\linebreak[4]
IRAS 01202+6133 is surrounded by an infrared nebula referred to as S187~IR \citep{Hodapp}.
\citet*{Zavagno} discovered an optically visible young stellar object, S187\,H$\alpha$.
The optical spectrum of this object exhibits several emission lines with P~Cygni type
profiles. It is probably a Herbig Ae/Be star.

The left panel of Fig.~\ref{Fig_S187_1} shows the large-scale distribution of the
molecular gas overlaid on the DSS image of the field \citep{Joncas}. The right
panel shows the map of CO outflow observed by \citet*{Casolib}, superimposed on $V$
frames of the S\,187 field obtained by \citet{Zavagno}. The pre-main sequence star
S\,187\,H$\alpha$ is indicated. Ammonia emission observed by \citet{Torrelles}
is shown by thick lines.

\begin{figure*}[tb]
\centerline{
\includegraphics[width=6cm]{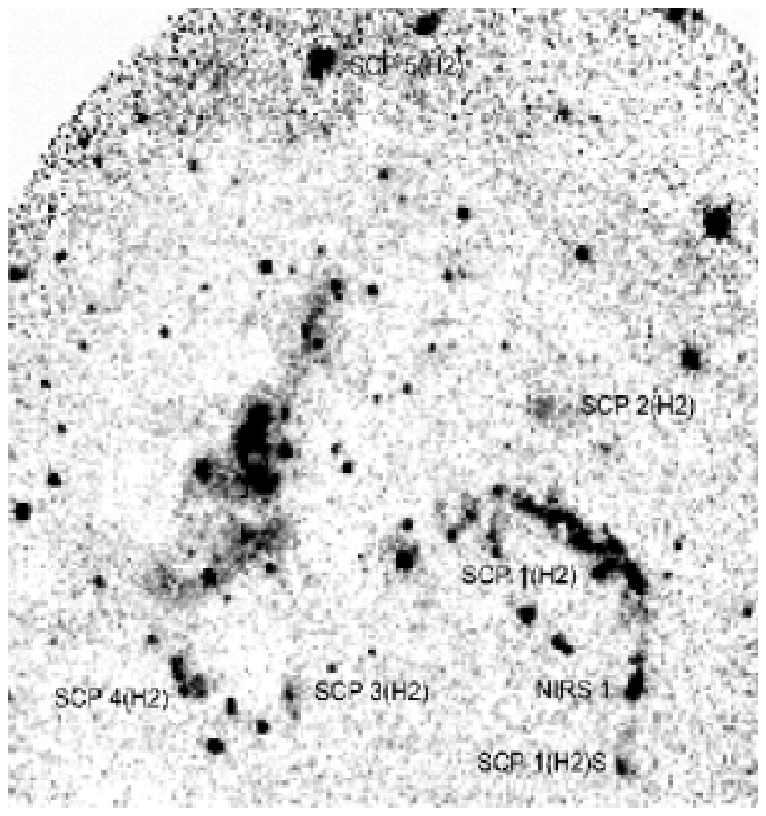}
\hskip2mm
\includegraphics[width=6cm]{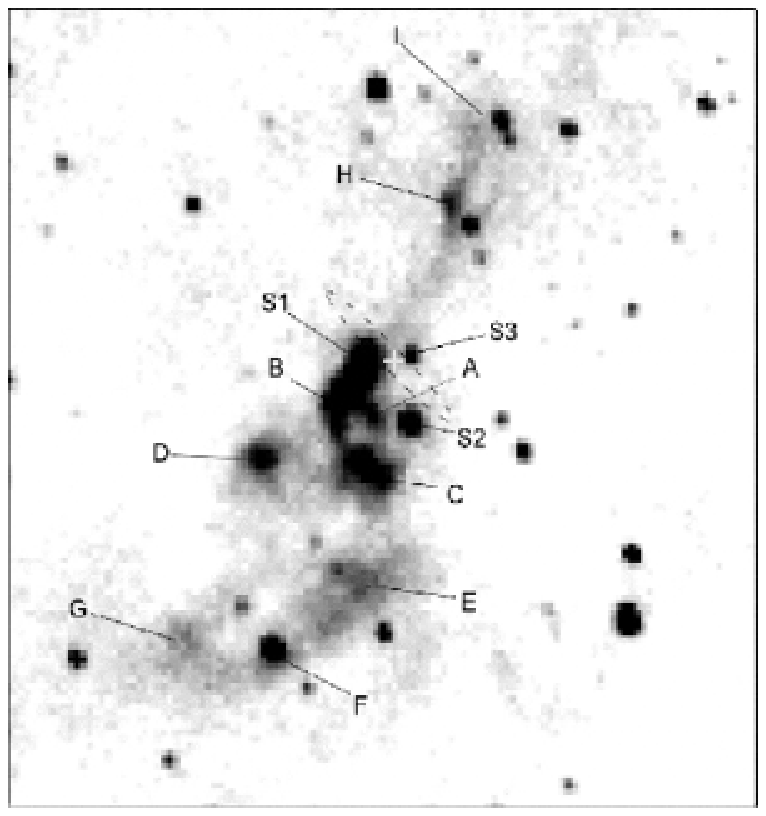}
}
\caption{{\it Left\/:} H$_{2}$ (2.122~$\mu$m) image of S187~IR in a
field of $3\farcm5 \times 3\farcm5$ (Fig. 1 of \citet{Jiang01}). The
H$_2$ knots detected by \citet{Salas} are labeled. A new H$_2$ knot in
the southwestern corner of the frame is labeled as SCP~1~(H2)\,S. The
curved H$_2$ jet, SCP~1 ~(H2), is significant to the southwest of the
S187~IR nebula, and three other H$_2$ knots are found to the north and
south, generally along the nebular extension. North is up, and east is
to the left.  {\it Right\/:} K' image of the bipolar infrared nebula
S\,187~IR, observed by \citet{Jiang01}.  The associated NIR point
sources (S1--S3) and reflection knots (A--I) are labeled. The dashed
ellipse indicates the uncertainty ellipse around the position of
IRAS~01202+6133.}
\label{Fig_s187_nir}
\end{figure*}

\citet*{Salas}  imaged in the near-infrared a $3\farcm67\times3\farcm67$ region
(corresponding to $1.07 \times 1.07$~pc$^2$), around IRAS 01202+6133.
They discovered a curved molecular hydrogen outflow that extends over a region of 76\arcsec \
(0.38\,pc at d=1\,kpc), and identified as S187:SCP~1\,(H2). The outflow changes
direction by more than 90\deg \ in a continuous way (see the left panel of  Fig.~\ref{Fig_s187_nir}).
The outflow-driving source is probably an extreme T Tauri star, identified as
S\,187~NIRS~1, located at the apex of the curved structure
(identical with 2MASS\,J\,01231828+6147388).
\citet{Salas} detected more than a hundred stars in the field.
They associate the IRAS point source  01202+6133 with the reddest source in the
field, with $K = 11.7$ and $(H-K) = 4.33$\,mag (not detected in J)
at the position RA(1950)=$01^\mathrm{h}20^\mathrm{m}16\fs1$ and
Dec(1950)=$+61\deg33\arcmin9\farcs8$, (identical with 2MASS\,J\,01233317+6148482)
which lies within 13\arcsec \ of the IRAS position.
This object could be a protostar or a highly extincted
intermediate-mass star ($L < 2800~L_{\sun}$).

\citet{Jiang01} present K'-band polarimetric images of S\,187~IR. They detected a bipolar,
near-infrared nebula around IRAS~01202+6133, with the southern part being bright
and knotty, and the northern part being faint and filamentary. The IRAS source 01202+6133 is
located at the center of the nebula and is associated with the reddest point source of the
field, labeled S1. The K'-band image  is shown in the right panel of Fig.~\ref{Fig_s187_nir}.
The polarization map of S\,187~IR reveals that the bipolar nebula is illuminated  by a single
central source.
\citet{Bica03} discovered an infrared cluster at $l=126\fdg66, b=-0\fdg79$, associated
with S\,187, using the data in the 2MASS All Sky Catalog (object No. 52).
S\,187 has been included in the  HCN (J=1-0) survey of bright infrared sources by \citet{Pirog99}
and in the N$_{2}$H$^{+}$(1-0) survey of massive molecular cloud cores by \citet{Pirog03}.

\section{Star Forming Regions beyond 2~kpc}

\subsection{MWC\,1080: Star formation in  L\,1238}

In spite of its large distance from the Sun, MWC~1080 (V628 Cas, HBC~317)
is among the most  thoroughly studied Herbig Be stars.
MWC~1080 was already proposed as a candidate early-type pre-main sequence star
by \citet{Herbig}.  It is embedded in the dark cloud L\,1238  at a
distance of 2.2-2.5 kpc \citep{Canto,Levreault}.
The spectral type, determined from the strength of the HeI lines by \citet{CK79},
is B0. From a low dispersion optical spectrogram \citet{Yoshida92} obtained
A0-A3. \citet{Hernan04} classified MWC\,1080 as a continuum star. The object is a
hierarchical multiple system: the primary, which is itself an
eclipsing binary \citep{Grankin,Leinert97} has an infrared companion separated by 0\farcs75
\citep{Leinert97} and another companion at 4\farcs7 \citep[e.g.][]{Pirzkal}.
\citet{Zinn94} detected strong X-ray emission from the object.
The whole system is surrounded by a small cluster of
infrared sources within a radius of 0.7\,pc \citep{Testi98}.
\citet{Canto} discovered a CO outflow from MWC\,1080.
\citet{MB01} detected  periodic polarimetric variations of MWC\,1080 at 7660~\AA.

\citet{Hillen92} classified MWC\,1080 on the basis of the shape of the SED
into group~I of the Herbig~Ae/Be stars. The SED of this group can be interpreted
by  a reprocessing circumstellar
disk (i.e. no accretion luminosity  is required) with a central hole.

\citet{Poetzel92} reported the discovery of Herbig-Haro outflow, HH~170, associated with MWC\,1080.
The HH emission was found to be predominantly east of the star, and a rather poorly
collimated flow is indicated. Long-slit spectroscopy of the
HH objects has shown that the radial velocities of the outflow are very high,
reaching a maximum value of 400~km/s in the line wings.

\citet{Yoshida91,Yoshida92} studied the optical spectrum of MWC\,1080, and map\-ped its
nearby nebulosity and the surrounding molecular cloud in several molecular lines.
They classified the spectral type as early A with a luminosity class of Ib-II.
They found  prominent P~Cygni profiles in the Balmer lines up to H$\epsilon$.

\citet{Abraham00} observed MWC\,1080 at mid- and
far-infrared wavelengths with ISOPHOT, the photometer on-board the {\it Infrared
Space Observatory\/}. They established that  MWC\,1080 has a peculiar SED, different
from any other Herbig~Be stars. With a steep flux density increase below
3.6\,\micron, the ISOPHOT and ground-based data outline a plateau in $F_\nu$ between 3.6 and
15\,\micron. Around 18\,\micron, however, the spectrum starts a second steep increase
which culminates in a well-defined peak at 100\,\micron.

The environment of MWC\,1080 was resolved by KAO at 100\,\micron. \linebreak \Citet{diFran94}
derived an apparent size of 29\arcsec.
The presence of an extended component is also inferred
from the submm observations of \citet{Mannings94}, who measured 350 and 450\,\micron \ flux
densities in excess to the prediction of a simple optically thick accretion disk
model. The source is also extended at 1.3~mm. Near-IR interferometric measurements
at 2.2\,\micron \  resolved the inner
($r \la 5\arcsec$) disk region of  MWC\,1080 \citep{Eisner03,Eisner04}. \citet{Testi98}
detected a conspicuous group of stars embedded in a diffuse nebulosity around MWC\,1080.
The radius of the group is about 0.7\,pc.

\citet{Wang07} present CS J=2-1, $^{13}$CO J=1-0, and C$^{18}$O J=1-0,
observations with the 10-element BIMA array toward the young
cluster around MWC\,1080. These observations reveal a biconical outflow
cavity with size $\sim$0.3 and 0.05 pc for the semimajor and semiminor axis and
$\sim$45\deg \  position angle. These transitions trace the dense gas, which is
likely the swept-up gas of the outflow cavity, rather than the remaining natal gas
or the outflow gas. 32 clumps are identified in the dense gas. The clumps
are approximately gravitationally bound, which suggests that they are likely
collapsing protostellar cores.

\subsection{The OB Associations in Cassiopeia}

\begin{table}
\caption{OB associations in Cassiopeia}
\label{Tab_ass}
\smallskip
\begin{center}
{\footnotesize
\begin{tabular}{lcccccccc}
\tableline
\noalign{\smallskip}
Name & $l_{min}$ & $l_{max}$ &  $b_{min}$ & $b_{max}$  & d\,(kpc) & \multicolumn{3}{c}{Number of massive stars} \\
     &          &            &            &            &          &  O type &  B type  & A type \\
\noalign{\smallskip}
\tableline
\noalign{\smallskip}
Cas OB\,2 & 110.1 & 114.0 & $-$1.3 & ~1.8 & 2630 & 2 & 12 & 2 \\
Cas OB\,5 & 114.9 & 118.0 & $-$2.4 & ~1.3 & 2500 & 5 & 10 & 2 \\
Cas OB\,4 & 119.0 & 121.6 & $-$2.1 & ~2.0 & 2280 & 5 & 12 & $\cdots$ \\
Cas OB\,14 & 119.7 & 121.1 & $-$1.3 & ~2.5 & 1100 & $\cdots$ & 3 & 1\\
Cas OB\,7 & 121.7 & 125.2 & $-$0.9 & ~2.6 & 2500 & 1 & 14 & 1 \\
Cas OB\,1 & 122.3 & 125.8 & $-$2.3 & $-$0.4 & 2500 & $\cdots$ & 5 & $\cdots$ \\
Cas OB\,8 & 129.2 & 129.7 & $-$1.5 & $-$0.2 & 2900 & 1 & 10 & 3 \\
Cas OB\,6 & 133.8 & 138.0 & $-$0.3 & ~3.0 & 2190 & 17 & 8 & $\cdots$ \\
\noalign{\smallskip}
\tableline
\end{tabular}}
\end{center}
\end{table}

A comprehensive study of the most luminous members of OB associations of the Cassiopeia
region beyond 2~kpc has been published by \citet{Humphreys}. Further results on membership
and Hertzsprung--Russell diagrams can be found in \citet{GS92}.
The basic data of the OB associations, taken from \citet{Humphreys}, are shown in
Table~\ref{Tab_ass}. All of them, except Cas~OB\,14, are located in the Perseus spiral arm.
The interactions of the high luminosity association members with their environments
are reflected by the giant far infrared loops, discovered in the IRAS images around
Cas~OB\,1, Cas~OB\,5, Cas~OB\,6, Cas~OB\,7, and Cas~OB\,8 by \citet{Kiss04}.
The shell around Cas~OB\,5 has been studied in detail by \citet{Moor}.

\acknowledgements This work was supported by the Hungarian OTKA grant\linebreak T049082.
I am grateful to JoAnn O'Linger for sending her results on L\,1340\,B before
publication, and to L\'aszl\'o Szabados for careful reading of the manuscript.
I used the {\it Simbad\/} and {\it ADS\/} data bases throughout this work.
Comments by the referee, Yoshinori Yonekura, are greatly appreciated.

\end{document}